\documentclass{aa}  

\usepackage{amsmath}	
\usepackage{amssymb}	

\usepackage{listings}
\usepackage{multirow}
\usepackage{graphicx}
\usepackage{txfonts}
\usepackage{color}
\usepackage{hyperref}
\begin{document}

   \title{Detectability and Characterisation of Strongly Lensed Supernova Lightcurves in the Zwicky Transient Facility}
   \titlerunning{glSN lightcurves in the ZTF}

   \author{A. Sagu\'es Carracedo\inst{1} 
    \and A. Goobar\inst{1} 
    \and E. M\"ortsell\inst{1}
    \and N. Arendse\inst{1} 
    \and J. Johansson\inst{1} 
    \and A. Townsend\inst{2} 
    \and S. Dhawan\inst{3} 
    \and J. Nordin\inst{2} 
    \and J. Sollerman\inst{4}
    \and S. Schulze\inst{5}  
    }

   \institute{The Oskar Klein Centre, Department of Physics, Stockholm University,
        Albanova University Center, SE 106 91 Stockholm, Sweden\\
              \email{ana.sagues-carracedo@fysik.su.se}
         \and
         Institut f\"ur Physik, Humboldt-Universit\"at zu Berlin, Newtonstr. 15, 12489 Berlin, Germany
         \and
         Institute of Astronomy and Kavli Institute for Cosmology, University of Cambridge, Madingley Road, Cambridge CB3 0HA, UK
         \and 
         Department of Astronomy, The Oskar Klein Center, Stockholm University, AlbaNova, 10691 Stockholm, Sweden
         \and
         Center for Interdisciplinary Exploration and Research in Astrophysics (CIERA), Northwestern University, 1800 Sherman Ave., Evanston, IL 60201, USA
             }
             
   \date{Received --- XX, XXXX; accepted --- XX, XXXX}

 
  \abstract
   {The Zwicky Transient Facility (ZTF) was expected to detect more than one strong gravitationally-lensed supernova (glSN) per year, but only one event was identified in the first four years of the survey.}
   {This work investigates selection biases in the search strategy that could explain the discrepancy and revise discovery predictions. }
   {We present simulations of realistic lightcurves for lensed thermonuclear (glSNIa) and core-collapse supernova (glCCSN) explosions over a span of 5.33 years of the survey, utilizing the actual observation logs of ZTF.  
}
   {We find that the magnitude limit in spectroscopic screening significantly biases the selection towards highly magnified glSNe, for which the detection rates are consistent with the identification of a single object by ZTF. To reach the higher predicted rate of detections requires an optimization of the identification criteria for fainter objects. We find that around 1.36 (3.08) SNe Ia (CCSNe) are identifiable with the magnification method per year in ZTF, but when applying the magnitude cut of $m<19$ mag, the detection rates decrease to 0.17 (0.32) per year.  We compare our simulations with the previously found lensed SNe Ia, iPTF16geu and SN Zwicky, and conclude that considering the bias towards highly magnified events, the findings are within expectations in terms of detection rates and lensing properties of the systems. In addition, we provide a set of selection cuts based on simple observables to distinguish glSNe from regular, unlensed, supernovae to select potential candidates for spectroscopic and high-spatial resolution follow-up campaigns. We find optimal cuts in observed colours  $g-r$, $g-i$, and $r-i$ as well as in the colour SALT2 fit parameter. The developed pipeline and the simulated lightcurves employed in this analysis can be found in the \texttt{LENSIT} github repository.
   }
   {}


   \maketitle
%

\section{Introduction}

When a galaxy lies in the line of sight of a background supernova explosion, the light from the supernova can be deflected and magnified by the gravitational field of the intervening galaxy. This effect is known as gravitational lensing, and in the regime of strong lensing we refer to these as gravitationally lensed supernovae (glSNe). In these cases, multiple images of the supernova are produced, each with different magnification and arrival time, depending on the geometry of the alignment and the properties of the galaxy acting as the lens.

glSNe provide excellent opportunities in cosmology and astrophysics  \citep[see][for a recent review]{2024SSRv..220...13S}. 
Interestingly, glSNe also uncover a population of compact galaxy-scale lenses which would not be possible to identify in the absence of a well-understood background source \citep{2023NatAs...7.1098G}. Compared to multiply-imaged quasars (QSOs), the main sources for time-delay cosmography today \citep{2016A&ARv..24...11T}, glSNe offer the advantage of shorter timescales, and they fade away, such that the lens galaxy can be studied without source contamination. They have a significantly larger amplitude of variability and a well understood family of light curves for accurate time-delay measurement. 
If the lensed source is a Type Ia SN, the small dispersion in the corrected peak luminosity helps to break modelling degeneracies and provides a magnification estimate independent of the mass model, and also yields robust time-delay estimates.

Identifying spatially unresolved glSNe is a challenging task. Optical sky surveys like PanSTARRS \citep{Kaiser2010}, the All\-Sky Automated Survey for Supernovae \citep[ASAS\-SN;][]{Shappee2014}, the Dark Energy Survey \citep[DES;][]{des}, the Asteroid Terrestrial \-impact Last Alert System  \citep[ATLAS;][]{Tonry2018} or the Zwicky Transient Facility \citep[ZTF;][]{Graham2019, Bellm2019PASP..131a8002B} detect thousands of supernova explosions every year. However, the probability of a massive galaxy being in the line of sight between the observer and the supernova explosion is very small. This probability increases with distance, as the number of galaxies along the path is statistically larger. Strong lensing can also occur from galaxy clusters; they have larger cross sections than individual galaxies but they are also much less abundant. In this paper, we will focus on galaxy-scale lensing systems.

In the past few decades, several glSNe have been discovered and studied, which have provided crucial insights into the properties of these events and their host galaxies. Five glSNe have been discovered behind galaxy clusters: SN Refsdal \citep{Kelly2015Sci...347.1123K} the first multiply- imaged glSN in a galaxy cluster, which allowed for a 6\% precision measurement of the Hubble constant, $H_0$, \citep{Grillo2018ApJ...860...94G, Kelly2023Sci...380.1322K}, 
SN ``Requiem'' \citep{Rodney2021NatAs...5.1118R}, AT2022riv \citep{Kelly2022TNSAN.169....1K}, ``C22''\citep{Chen2022Natur.611..256C}, ``SN H0pe'' \citep{Frye2023TNSAN..96....1F} and SN Encore \citep{Pierel_2023_Encore}. Two glSNe have been discovered behind galaxy-scale lenses: SN iPTF16geu \citep{Goobar2017Sci...356..291G}, the first multiply- imaged glSN by a single-galaxy discovered with the intermediate Palomar Transient Factory \citep[iPTF;][]{Kulkarni2013ATel.4807....1K}, predecessor of ZTF, and SN Zwicky \citep{2023NatAs...7.1098G}, both of them produced by thermonuclear explosions. The latter being the one glSN found with the ZTF so far. There are two main identification methods: 
either they are differentiated by their magnified brightness (``magnification method'') or they are clearly resolved as multiple images of the same source (``multiplicity method''). In the single-galaxy lensing systems that we focus on in this paper, the image separation of the multiple images is too small to be resolved with current ground-based surveys. 

Finding single-galaxy-lensed supernovae can greatly enhance our understanding of the properties of small lensing galaxies, measure the expansion rate of the universe \citep{Refsdal1964MNRAS.128..307R} and provide constraints on other cosmological parameters \citep{2002A&A...393...25G}. Upcoming surveys, such as the Vera Rubin Observatory's Legacy Survey of Space and Time \citep[LSST;][]{Ivezi2019ApJ...873..111I} and the Nancy Grace Roman Space Telescope \citep[Roman;][]{Pierel2021ApJ...908..190P}, are expected to discover tens to hundreds of glSNe per year \citep{Arendse2023arXiv231204621A, Sainz2023MNRAS.526.4296S, Wojtak2019MNRAS.487.3342W, Goldstein2017ApJ...834L...5G, Oguri2010MNRAS.405.2579O}. 
It is important  to gain understanding on the observational properties of glSN lightcurves and improve detection methods to improve the detection rates for glSNe. 

According to \cite{Goldstein2019ApJS..243....6G}, it was expected that ZTF would detect more than one glSN per year. However, after four years of ZTF's operations, only one glSN, SN Zwicky, was identified. The observational challenges and sub-optimal search criteria might explain this discrepancy. In this study, we explore selection effects affecting the identification of all types of glSNe. SN Zwicky and iPTF16geu, the latter found before ZTF, were identified with the magnification method, because they were bright enough to obtain a classification spectrum. Other glSNe might be in the ZTF database but never passed this magnitude limit for spectroscopic vetting and are unidentified or misidentified ZTF transients. In a parallel paper, Townsend et al. (2024 in prep.) perform an archival search for glSNe with ZTF, which utilizes results from this paper to improve the selection criteria, we will refer to this paper as T24. 

The primary objective of this study is to explore the characteristics of synthetic lightcurve data, simulated according to the real observations of ZTF, 
to investigate the selection criteria require to identify glSN within ZTF. We simulate realistic unresolved glSN lightcurves using the actual observing logs of the ZTF survey and analyze how the expected detection rates from idealized survey predictions compare to the actual observing performance. We determine the number of glSNe as a function of magnitude limit, and explore the observables that can be used for background rejection.

Section~\ref{sec:simulations} explains the simulation packages, the modeling assumptions we make on the glSN population used in this analysis, and the methodology used to produce synthetic lightcurves. 
Section~\ref{sec:detecting_glsne} describes the detection criteria used and investigates detectability and selection cuts. In Sect.~\ref{sec:analysis method} we analyse the lensing parameters obtained from the synthetic lightcurves and compare to SN Zwicky and iPTF16geu.
Finally, in Sect.~\ref{sec:conclusion} we discuss our findings, the likelihood of finding glSNe such as SN Zwicky and iPTF16geu, propose an explanation for the discrepancy between predicted number and actual number of glSN detections, and discuss our actual performance and improvement and prospects on finding more lensed SNe.

\section{Simulations}\label{sec:simulations}

\subsection{Strong lensing simulation}\label{sec:lensing_sim}

The probability of a supernova to be subjected to strong lensing involves the lensing cross-section, which depends on the velocity dispersion of the lens galaxies, and the co-moving volume as follows \citep{Oguri2019RPPh...82l6901O}: 
\begin{equation}\label{eq:Psl}
    P_{sl}(z_s) = \int^{z_{\rm s}}_0 \rm{d}z_l \frac{\rm{d}^2V }{\rm{d}z_l\rm{d}\Omega} \int^{\infty}_0 \rm{d}\sigma \frac{\rm{d}n}{\rm{d}\sigma} B A_{sl}(\sigma),
\end{equation} where $z_{\rm s}$ is the redshift of the source, $z_{\rm l}$ the redshift of the lens galaxy, $\frac{\rm{d}^2V }{\rm{d}z_l\rm{d}\Omega}$ the comoving volume element per redshift and steradian, $\frac{\rm{d}n}{\rm{d}\sigma}$ the velocity dispersion function of galaxies, $B$ the magnification bias and $\rm A_{sl}(\sigma)$ the lensing cross-section dependent on the velocity dispersion ($\sigma$). 

The input sample in our simulation for a given glSN type would follow a redshift distribution that combines the volumetric supernova rate with the lensing probability described above \citep{Oguri2019RPPh...82l6901O},
\begin{equation}\label{eq:R_sl}
    R_{sl}(<z_{max}) = \Omega_{\rm sky} \int^{z_{\rm max}}_0 \rm{d}z_s\frac{\rm{d}^2V }{\rm{d}z_s\rm{d}\Omega} \frac{R(z_s)}{1+z_s} P_{sl},
\end{equation} with $R(z_{\rm s})$ the redshift dependent volumetric rate of the supernova type. $\Omega_{\rm sky}$ corresponds to the area of the sky probed. In the case of ZTF we simulate around 75\% of the whole sky.

We assume a Singular Isothermal Ellipsoid \citep[SIE;][]{Kormann1994A&A...284..285K} model with external shear for the lens galaxies.

The redshift of a lens galaxy is sampled from:
\begin{equation}\label{eq:z_l}
    \rm P(z_l) = \frac{1}{K} \frac{(1+z_l)^2 D_l^2}{E(z_l)}\ ,
\end{equation} with  $ K = \int^{z_{\rm l,max}}_{z_{\rm l,min}} \frac{(1+z_{\rm l})^2 D_l^2}{E(z_{\rm l})} dz_l$ and $D_l$ the angular diameter distance of the lens galaxy. $E(z_l) = \sqrt{\Omega_m (1+z_l)^3 + (1-\Omega_m)} $ is the assumed cosmology considered from the latest results from the Planck mission \citep{Planck2020A&A...641A...6P}. We thus
 assume a flat $\Lambda$CDM model with $H_0=67.8$ km  s$^{-1}$ Mpc$^{-1}$ and $\Omega_m=0.308$. 

The velocity dispersion ($\sigma$) is modeled as a modified Schechter function \citep{Sheth2003ApJ...594..225S}:
\begin{equation}\label{eq:sigma}
    \frac{\rm{d}n}{\rm{d}\sigma} = \phi_{*} \left(\frac{\sigma}{\sigma_{*}}\right)^{\alpha} \exp\left[-\left(\frac{\sigma}{\sigma_{*}}\right)^\beta\right]\frac{\beta}{\Gamma(\alpha/\beta)}\frac{d\sigma}{\sigma},
\end{equation} with parameter values $\alpha = 2.32$, $\beta= 2.67$, $\phi_{*} = 8\times 10^{-3} \rm \ h^3 \ Mpc^{-3}$ and $\sigma_{*}= 161 \rm \ km \ s^{-1}$, internal galaxy properties inferred from Sloan Digital Sky Survey (SDSS) in \cite{Choi2007ApJ...658..884C}. 

The Einstein radius for a SIE model is given by:
\begin{equation}\label{eq:einstein_radius}
    \theta_E = 4\pi \left( \frac{\sigma}{c}\right)^2 \frac{D_{ls}}{D_s},
\end{equation} where $D_{ls}$ and $D_s$ are the angular diameter distance between the lens and the source and between the observer and the source, respectively, and $\sigma$ is the velocity dispersion and $c$ the speed of light. 

The ellipticity ($e$) is distributed following a velocity dispersion-dependent Rayleigh density as described in \citet{Collett2015ApJ...811...20C}
\begin{equation}\label{eq:ellipticity}
    \rm P(e|\sigma) = \frac{e}{s^2}\exp \left(-\frac{e^2}{2s^2}\right),
\end{equation} with $s=A+B\sigma$ with fit values  $A = 0.38$ and $B = 5.7\times10^{-4}$ (km s$^{-1}$)$^{-1}$. To avoid including extremely elongated mass profiles, the ellipticity parameter is limited to a maximum value of $e=0.8$. We assume random orientation for the lens and the external shear: $\theta_e \sim U[0,2\pi]$, and $\theta_{\gamma}\sim U [0,2\pi]$. We assume a Rayleigh distribution in magnitude for the external shear: 
\begin{equation}\label{eq:gamma}
    P(\gamma|s) = \frac{\gamma}{s^2} \exp\left(- \frac{\gamma^2}{2s^2}\right),
\end{equation} with $s=0.05$  \citep{Wong2011ApJ...726...84W}.

\begin{figure}
    \centering
    \includegraphics[width=\linewidth]{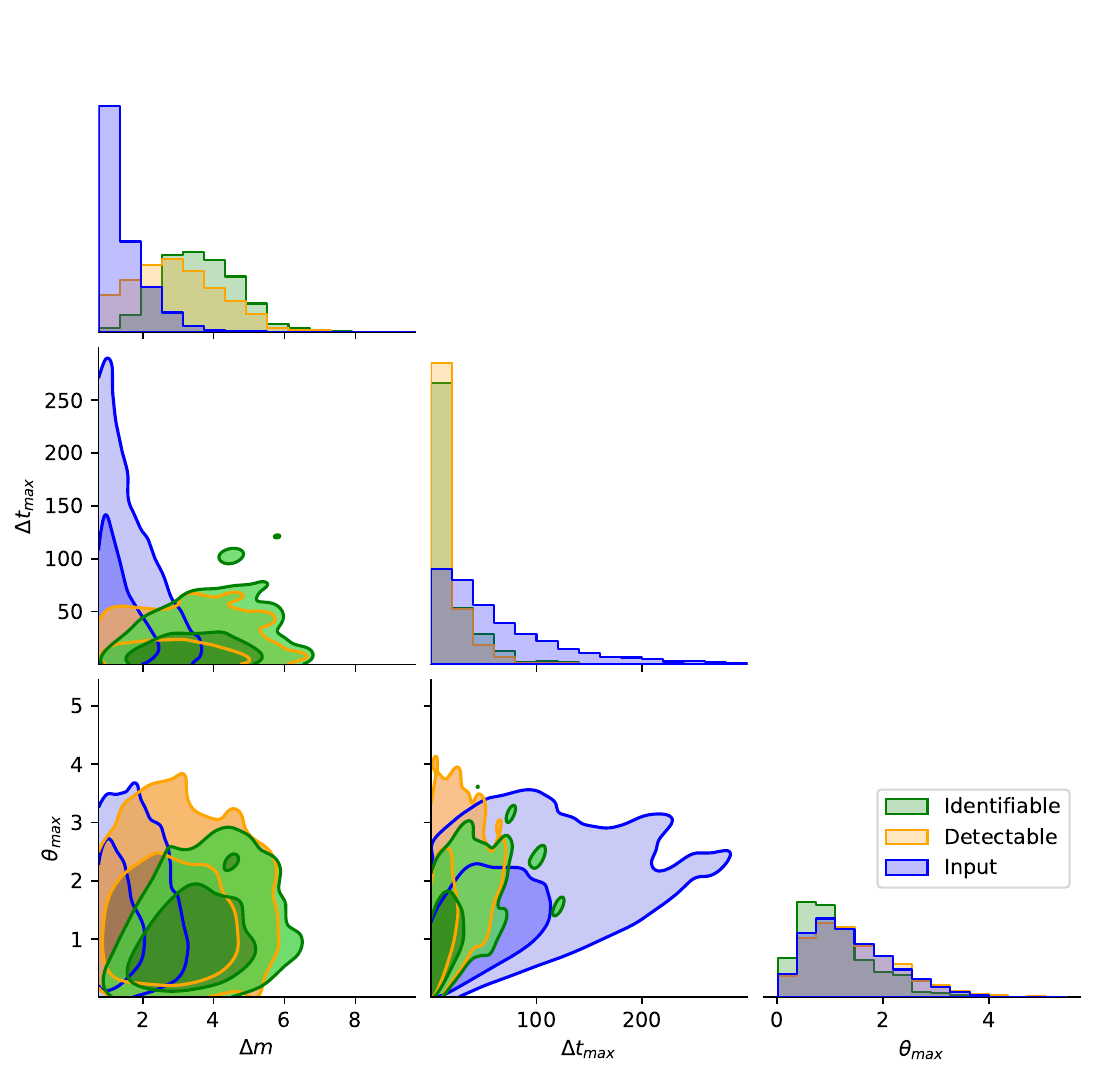}
    \caption{Distribution of lensing parameters generated for a population of Type Ia supernovae up to redshift 1.5 under the assumptions described in Sect.~\ref{sec:lensing_sim}. The distributions are separated into input data (blue), the detectable sample (orange) and the identifiable sample (green). The detectable criteria correspond to basic requirements on the lightcurve (minimum 5 detections around the peak) and identifiable refer to those that in addition pass the magnification method (see Sect.~\ref{sec:detection criteria}).
    The lensing parameters shown in this plot corresponds to the total magnification in magnitude ($\Delta m$), the maximum time delay ($\Delta t_{max}$) and the maximum angular separation in arcsecond ($\theta_{\rm max}$). }
    \label{fig:input_lensparams}
\end{figure}

The source position is uniformly distributed in an area within the Einstein radius in polar coordinates: $r \sim U[0,1]$, $\theta \sim U[0,2\pi]$. The source positions are transformed to cartesian coordinates by:
$x_s = \theta  \sqrt{r} \cos(\theta)$ and $y_s = \theta  \sqrt{r} \sin(\theta)$, both in angular units. 

We simulate source redshifts up to $z_{\rm s, max}=1.5$. We utilize the software package \texttt{lenstronomy}\footnote{\url{https://lenstronomy.readthedocs.io/en/latest/} }\citep{Birrer2018PDU....22..189B, Birrer2021JOSS....6.3283B} to solve the lensing equation with parameters sampled following the formulae described in this section. We provide source position, redshifts, parameters for the lens model, and after running \texttt{lenstronomy} we obtain the multiplicity, image positions, magnification ($\mu_i$) and arrival time ($t_i$) for each image. 
We iterate until each $z_s$ get a set of parameters that satisfies multiplicity ($N_{\rm im}\geq2$) and minimum total magnification $\mu_{\rm total}>2$, where the total magnification is calculated by the sum of the individual magnifications: 
\begin{equation}
    \mu_{\rm total} = \sum_{i}^{N_{\rm im}} \mu_i
\end{equation}

The angular separation between two images $i$ and $j$ in the system is calculated by: 
\begin{equation}
    \theta_{i,j} = \sqrt{ (x_i-x_j)^2 + (y_i-y_j)^2 },
\end{equation}with $x_i$ and $y_i$ the image positions in cartesian coordinates, with $i,j$ from 1 to the number of multiple images.

The resulting distribution of parameters for the magnification, arrival time and image separation from our assumptions are shown in Fig.~\ref{fig:input_lensparams}.

With \texttt{lenstronomy} and the assumptions on the lensing model described in this  section, we obtain 3 types of multiplicities: Doubles (two images), Triplets (three images) and Quads (four images). The relative proportions on the input sample are $\sim$87\% doubles, $\sim$1\% triplets and $\sim$12\% quads. We observe that the most common setting provides doubles, but they correspond to smaller total magnifications. Triplets are rare but still possible, they typically correspond to higher ellipticities and shear. In the detectable sample, the relative multiplicity fractions become $\sim$32\% doubles, $\sim$3\% triplets and $\sim$65\% quads. A more detailed discussion about the detected multiplicity is given in Sect.~\ref{subsec:multiplicity}.

\subsection{Modeling unresolved glSN lightcurves}\label{sec:lcsmodel}

The angular separations between multiple images are expected to be small compared to the spatial resolution of around 2 arcseconds for ZTF. Therefore, it is fair to assume that the majority (if not all) of the glSNe will have unresolved lightcurves in ZTF. 

To model the unresolved lightcurves we have created a source class wrap that takes any source class implemented in the python library for supernova cosmology \texttt{SNCosmo}  \footnote{\url{https://sncosmo.readthedocs.io/en/stable/}} \citep{Barbary2016ascl.soft11017B} and applies the strong lensing effect to derive the multiplicity ($N_{\rm im}$), time delays (time arrival differences: $dt_i$), and flux magnifications ($\mu_i$). This model allows for including additional  effects on the lightcurves, such as the extinction by dust in the host and lens galaxies and the Milky Way, as well as microlensing by stars in the lensing galaxy. We consider the standard extinction law for the wavelength dependence from \cite{Cardelli1989}, and Milky Way Galactic extinction based on the Schlegel, Finkbeiner, Davis (SFD98) reddening maps \citep{Schlegel1998ApJ...500..525S}. For the host galaxy extinction, we adopt a total-to-selective extinction ratio, $R_V=2$ and a colour excess described by an exponential function \citep[see e.g.,][]{Stanishev2018} with scale parameter $\beta=0.11$, for the Milky Way we adopt $R_V=3.1$. In this work, we assume no reddening in the lensing galaxy as it is unclear how to model it, and since the two glSNe found in single-galaxy lensing had negligible extinction \citep{2023NatAs...7.1098G, Goobar2017Sci...356..291G}, we also neglect the effects of microlensing in the model.
Dust extinction and microlensing in the lens galaxy have opposites effects in the observed brightness of the SN, with the latter possibly adding on extra magnification.

For unresolved lightcurves, the total flux as function of time consists of the addition of the fluxes of each SN image at their arrival times ($t_i$) multiplied by their magnifications ($\mu_i$).

\begin{equation}
    F_T = \sum_i^{N_{\rm im}} F(t_i) \mu_i
\end{equation}

We simulate thermonuclear and core-collapse supernovae by utilising spectral energy distribution templates available in \texttt{SNCosmo}.
The \texttt{SNCosmo} template name is an input for our model as well the corresponding intrinsic parameters of the given template and the lensing parameters including the arrival times difference with respect to image 1 ($dt_i$) and the magnification of each individual image. 

The simulated supernovae  are uniformly distributed in right ascension (RA) and the declination (Dec) within the range covered by the ZTF survey: Between $0^{\circ}$ and $360^{\circ}$ in RA and from $-30^{\circ}$ to $+90^{\circ}$ in Dec. We assume homogeneity in the sky distribution for lensed events. Time of explosion is also considered homogeneous and we use a uniform distribution for this parameter in the time range of the observations. 

The amplitude parameter that the model requires is that of the supernova magnitude without magnification ($\rm M_{int}$). Then the absolute magnitude with lensing ($\rm M_{len}$) is computed as 
\begin{equation}
    \rm M_{len} = M_{int} - 2.5\cdot \log_{10}(\mu_{total})
\end{equation}

\begin{figure}
    \centering
    \includegraphics[width=\linewidth]{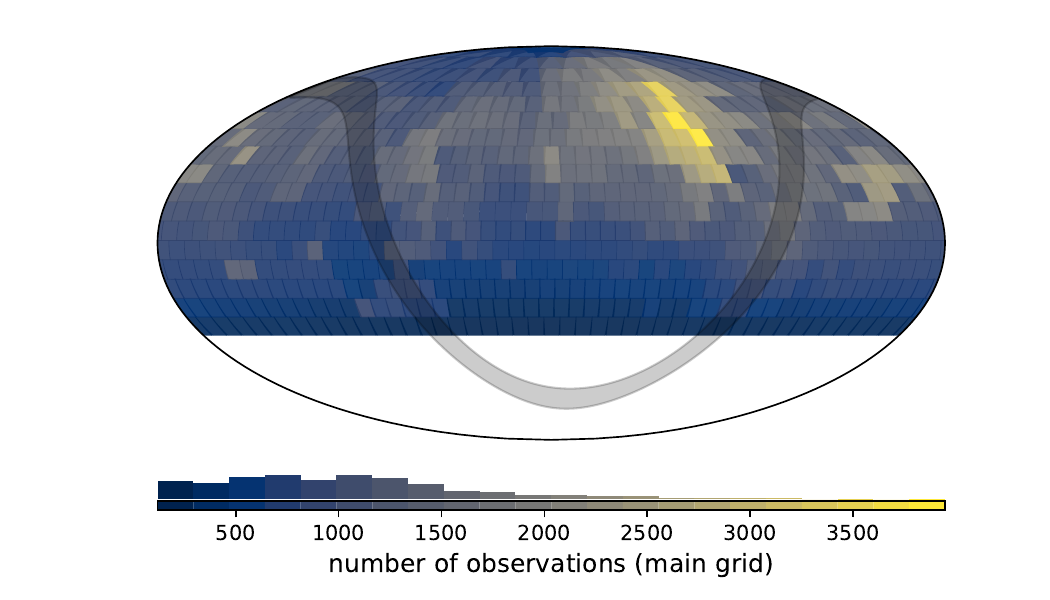}
    \caption{ZTF survey tiles coverage, including public survey and Partnership time. We see that most of the tiles had received around 1000 visits over its survey time, although some other fields received higher cadence visits, accumulating more than 3000 in total.}
    \label{fig:survey_show}
\end{figure}

\subsubsection{Modelling SNe Ia}

To simulate SNe Ia, we adopt SALT2 \citep{Guy2007A&A...466...11G}, an empirical model that parametrises the SN lightcurves in terms of stretch ($x_1$) and colour (c). We make use of SALT2 parameter distribution functions built into the \texttt{skysurvey}\footnote{\url{https://skysurvey.readthedocs.io/en/latest/}} package. The stretch parameter assumes that the population is described by two Gaussian distributions with mean values  $0.33$ and  $\-1.50$ and with standard deviations $0.64$ and $0.58$, respectively, and a prompt fraction of 0.5 is assumed \citep{Nicolas2021A&A...649A..74N}. For the colour parameter c, we assume an exponential decay distribution convolved with an intrinsic Gaussian with mean $c_{\rm int} = -0.075$ and $\sigma_{\rm c,int} = 0.05$. The intrinsic absolute magnitude is obtained by making use of the two parameter luminosity correlation from \cite{Tripp1998A&A...331..815T} $M =  M_0 + \alpha x_1 + \beta c$, with $\alpha = -0.14$, $\beta= 3.15$ and $M_0$ the absolute magnitude corresponding to $x_1=0$ and $c=0$ for a normal distribution is assumed with average $-19.3$ mag and an intrinsic scatter of $\sigma_{int} = 0.1$ mag. We assume SN Ia volumetric rates described in \cite{Kessler2019PASP..131i4501K} with local rate $\rm R_{loc} = 2.35\times 10^4$ Gpc$^{-3}$ yr$^{-1}$ and a redshift evolution described by $(z+1)^{1.5}$ for $z<1$ \citep{Dilday2008ApJ...682..262D} and  $(z+1)^{-0.5}$ for $z>1$  \citep{Hounsell2018ApJ...867...23H}.

\subsubsection{Modelling CC SNe}

We consider core-collapse supernova types IIP, IIn and Ibc. 
To simulate them we use the templates built in \texttt{SNCosmo} as in \cite{Goldstein2019ApJS..243....6G}, in particular we employ ``s11-2005lc'' template for IIP \citep{Sako2011ApJ...738..162S}, ``nugent-sn2n'' for IIn \citep{GillilNugentPhillipsand1999ApJ...521...30G} and ``nugent-sn2n'' for Ibc \citep{LevanNugent2005ApJ...624..880L}. The intrinsic magnitude is assumed to be distributed from a normal distribution in the rest-frame $B$-band absolute magnitude in the Vega system as described in Table \ref{tab:input_param}, with peak magnitude centered around -16.9 mag for IIP's, -19.05 for IIn's and -17.51 for Ibc. We choose to focus in these three CCSN subtypes as they are expected to contribute the most in detection rates from predictions in \cite{Goldstein2019ApJS..243....6G}. 

The local volumetric rates are taken from \cite{Perley2020ApJ...904...35P} and the redshift dependence follow the star-formation density as described in \cite{Madau2014ARA&A..52..415M}.

\subsection{Survey observation logs for ZTF}\label{sec:survey}

\begin{figure*}
    \centering
    \includegraphics[width=.49\linewidth]{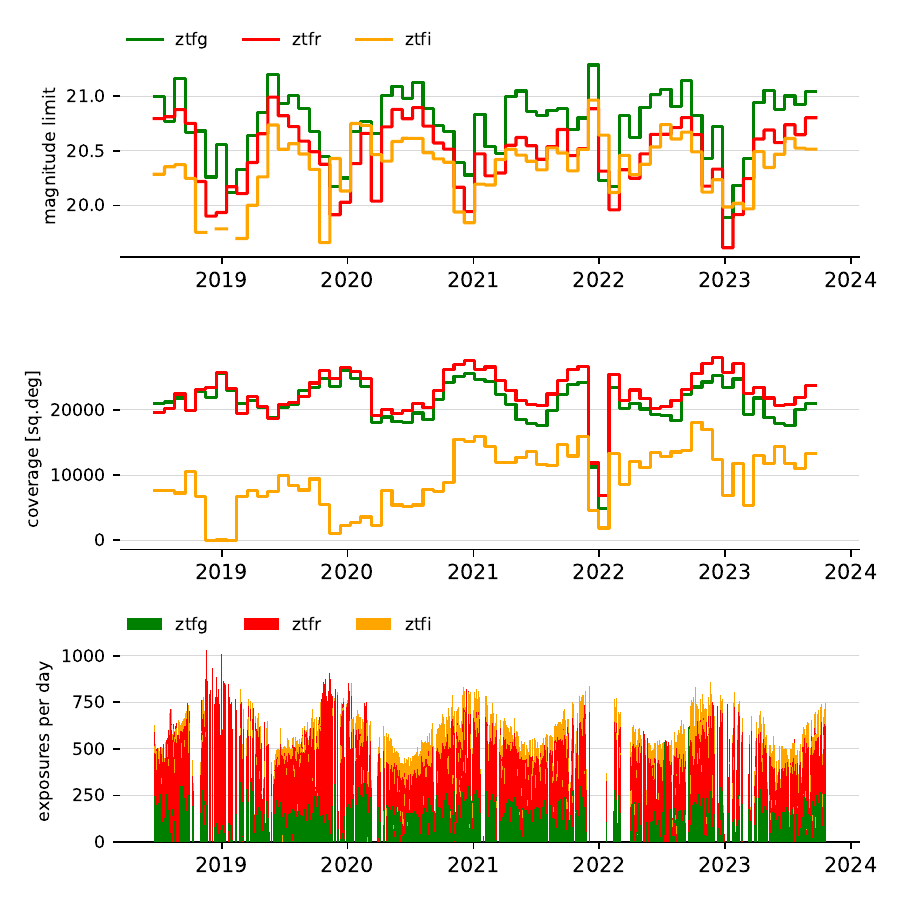}
    \includegraphics[width=.49\linewidth]{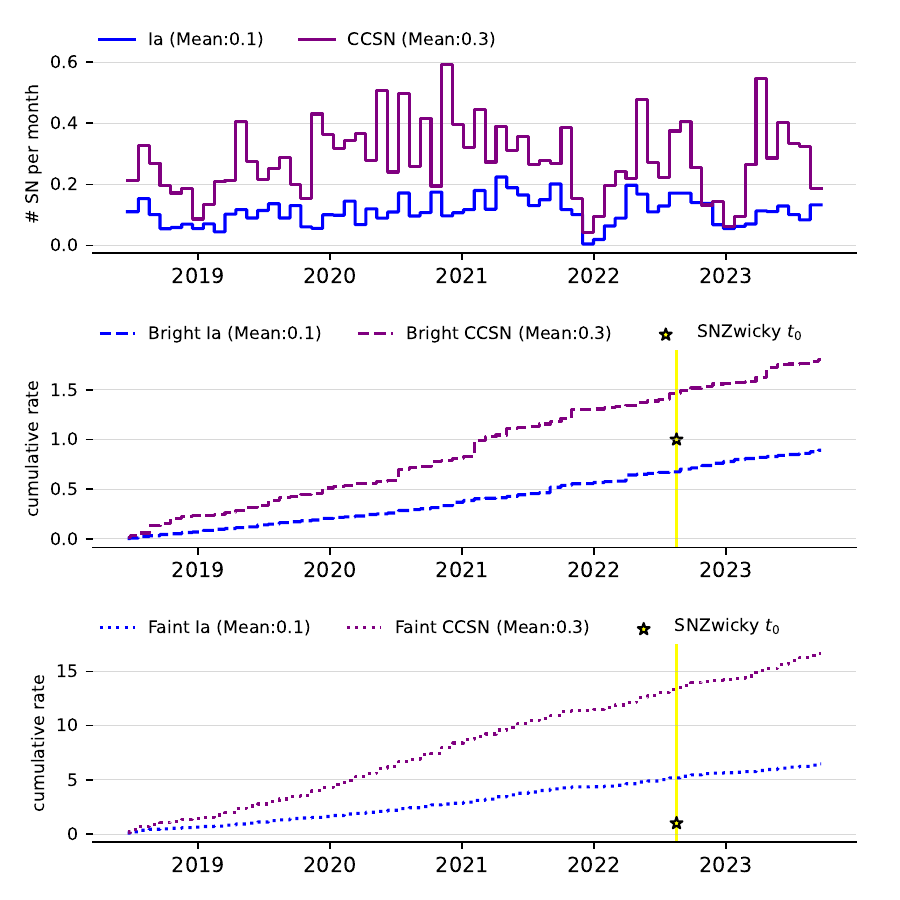}
    \caption{\textbf{Left panels.} This plot illustrates some of the key aspects that describes the performance of the ZTF along its survey time. Top panel: The fluctuations in depth of each filter ($g$, $r$ and $i$). Note the sinusoidal shape that correlates with the season, i.e. it is typically harder to achieve deeper limiting magnitudes in winter as compared to summer. We can also appreciate that the $i$-band depth improved over time. Middle panel: the squared degrees covered per month with each filter. We again observe a sinusoidal fluctuation and that $i$-band sky coverage increases over the survey time. The last thing is crucial, as redder bands are beneficial in the search for the redshifted lightcurves of lensed SNe. Bottom panel: Exposures per night and filter, showing a clear correlation with the length of the nights throughout seasons. Our log includes gaps in the observations, for example, due to bad weather or instrument issues. \textbf{Right panels.} The top panel shows the number of expected glSNe in the identifiable sample per month bins for SNe Ia (blue) and CCSNe (purple). The average monthly rate is 0.1 glSN-Ia and 0.3 glSN-CC. The middle and bottom panels show the cumulative detections per time, middle panel for the Bright sample ($m<19$ mag) and bottom panel for the Faint sample ($m>19$ mag). The yellow line indicates the time of peak of the first ZTF lensed Ia discovery, SN Zwicky. Finding $\sim1$ lensed SN Ia in 5 years in the Bright survey is consistent with our simulations.}
    \label{fig:survey_pointings}
\end{figure*}

The Zwicky Transient Facility project uses a camera mounted on the Samuel Oschin Telescope at the Palomar Observatory in California to scan the night sky for transient and variable objects \citep{Bellm2019PASP..131a8002B}. ZTF is led by the California Institute of Technology (Caltech) in collaboration with partners worldwide. It is a valuable facility for discovering new transients due to the wide field of view of 47 degrees squared and relatively high cadence of $\sim$3 days. 

ZTF uses three filters ($g$, $r$, and $i$) to map the northern sky to minimum declination of $-30^{\circ}$ (see Fig. \ref{fig:survey_show}). The $g$ and $r$ bands are used more frequently, while the $i$-band is over a smaller region of the sky. However, the second half of the survey includes more $i$-band observations, which is beneficial for the search for glSNe that appear redder than unlensed events mainly due to their higher redshifts. 
The survey operates year-round, and the number of visible fields and the length of the nights change throughout the year. 

For this work we are considering observations taken under the public survey, as well as initially proprietary "Partnership time", during 5.33 years of the survey from June 19th, 2018 to October 19th, 2023. The typical exposure time is 30 seconds, and the median depth of the survey is about 20.5 mag. However, this is not the same for all filters and varies over time and direction mainly due to atmospheric conditions, exposure lengths and detector sensitivity. 

Figure \ref{fig:survey_pointings} illustrates the survey performance from the observation logs used in this work. We note that the number of exposures, solid angle coverage and depth varies with the time of the year. During winter, dark nights are longer, hence more exposures. However, on average, the atmospheric conditions are worse, leading to shallower monthly depth average compared to summer nights, which explains the anticorrelated sinusoidal curves. Figure \ref{fig:survey_pointings}  also illustrates how the detection rate depends on the survey performance, by shown the number of detections over survey time. We note drops in detections correlated to drop in magnitude depth, and we appreciate a positive effect in the number of detections with the increase of $i$-band usage.

\subsection{Simulating synthetic lightcurves}

In order to generate realistic lightcurve observations, the Python package \texttt{skysurvey} is utilized. \texttt{skysurvey} is an open source package built in Python for astronomical survey simulations that has a better speed performance compared to its predecessor \texttt{simsurvey} \citep{2019JCAP...10..005F}. The simulation process involves the combination of a modeled population of glSNe with actual observing logs from the ZTF to obtain lightcurves as they would have been observed by the survey. This includes the replication of realistic error bars, cadence, and wavelength coverage.

The unresolved supernova lightcurve model, described in Sect.~\ref{sec:lcsmodel}, is provided. The lensing parameters obtained as described in Sect.~\ref{sec:simulations} are used as input parameters. The intrinsic supernova parameters are sampled independently of the lensing properties following the assumptions described in Sect.~\ref{sec:lcsmodel}.

The modeled lightcurves are combined with the observation log of ZTF described in Sect.~\ref{sec:survey}. Subsequently, \textit{skysurvey} is employed to obtain the observed lightcurves with realistic error bars and cadences.

\subsection{Simulation framework}

We develop a pipeline named \textit{Lensing End-to-end Supernovae Lightcurve Investigation Tool} (\texttt{LENSIT}\footnote{\url{https://github.com/asaguescar/LENSIT}}) that consists of a framework to combine \texttt{lenstronomy} and \texttt{skysurvey} to perform an end-to-end analysis of simulated unresolved lightcurves of supernovae with ZTF observing logs. The catalogs of synthetic data produced in this work are available in the github repository for \texttt{LENSIT}.

The first step is to run \texttt{lenstronomy} with the model assumptions described in Sect.~\ref{sec:lensing_sim} from which we obtain the relevant lensing parameters to add together with the supernova parameters in the unresolved glSN lightcurve model described in Sect.~\ref{sec:lcsmodel}. We run \texttt{skysurvey} with the lightcurve model combined with the ZTF observations described in Sect.~\ref{sec:survey} to simulate the realistic lightcurves. This lightcurves include realistic sampling, depths and cadences as if observed with ZTF. 

In this work we simulate a fixed number of supernovae in the order of $10^5$ strongly lensed and detected events that provides enough statistics for our analysis. To estimate a realistic number of events, we re-scale with the expected rates by using equation~\eqref{eq:R_sl}. The input parameters assumed in this work are summarize in Table~\ref{tab:input_param}.

\begin{table}
    \centering
    \caption{Input parameters for \texttt{lenstronomy} and for intrinsic supernova parameters. $\mathcal{N}(\mu,\sigma)$ represents gaussian distribution with mean $\mu$ and standard deviation $\sigma$. $\mathcal{E}(\rm loc,scale)$ is a exponential distribution shifted and scaled by the parameters $loc$ and $scale$. $\mathcal{U}(\rm min,max)$ corresponds to an uniform distribution between $min$ and $max$.}
    \label{tab:input_param}
    
    \begin{tabular}{llc}
        \hline\hline
        Parameter & Value or Distribution \\
        \hline
        \\
        \multicolumn{2}{p{.9\linewidth}}{Lens Galaxy Model} \\ \hline
            Lens model & "SIE" + "SHEAR" & \\
            Lens redshift &  $0<z_{\rm l}<z_{\rm s}$ \\
            Velocity dispersion & $50<\sigma<400$ \\
            Ellipticity & $0<e<0.8$ \\
            External shear & $0<\gamma<1.5$ \\
            Source position & $x_s, y_s \sim  \mathcal{U}(-\theta_E,\theta_E)$ \\
            Orientation angle (rad) & $\theta_{e}, \theta_{\gamma}\sim  \mathcal{U}(0,2\pi)$ \\
        \\
        \multicolumn{2}{p{.9\linewidth}}{Type Ia SN}
        \\
        \hline
            template & SALT2 \\
            Stretch & $x_1 \sim 0.7\mathcal{N}(0.33,0.64)$ \\
              & \ \ \ \ \ $+0.3\mathcal{N}(-1.5,0.58)$ \\
            Colour &     $c \sim \mathcal{N}(-0.075,0.05)$ \\
              &  \ \ \ \ \   $\times \mathcal{E}(-0.075, 0.05)$ \\
            Corrected magnitude & $M_0 \sim \mathcal{N}(-19.3,0.10)$ \\
            $\alpha$, $\beta$ & $-0.14$, $3.15$ \\
            Local rate & $\rm R_{loc} = 2.35\cdot 10^4 \ \rm Gpc^{-3} yr^{-1}$ \\
        \\
        \multicolumn{2}{p{.9\linewidth}}{Type IIP SN}
        \\
        \hline
            template & "s11-2005lc" \\
            Absolute magnitude & $ M_B \sim \mathcal{N}(-16.9,1.12)$ \\
            Local rate & $\rm R_{loc} = 5.52\cdot 10^4 \ \rm Gpc^{-3} yr^{-1}$  \\
        \\
        \multicolumn{2}{p{.9\linewidth}}{Type IIn SN}
        \\
        \hline
            template & "nugent-sn2n" \\
            Absolute magnitude & $M_B \sim \mathcal{N}(-19.05,0.5)$ \\
            Local rate & $\rm R_{loc} = 5.05\cdot 10^3 \ \rm Gpc^{-3} yr^{-1}$  \\
        \\
        \multicolumn{2}{p{.9\linewidth}}{Type Ibc SN}
        \\
        \hline
            template & "nugent-sn1bc" \\
            Absolute magnitude &  $M_B \sim \mathcal{N}(-17.51,0.74)$  \\
            Local rate & $\rm R_{loc} = 3.33\cdot 10^4 \ \rm Gpc^{-3} yr^{-1}$  \\
        \\
        \multicolumn{2}{p{.9\linewidth}}{Sky localization}
        \\
        \hline
            Right ascension ($^{\circ}$) &   $0<\rm RA<360$ \\
            Declination ($^{\circ}$)   &   $-30<\rm Dec<90$ \\
            Time (mjd)   &   $t_0 \sim \mathcal{U}(58288.2, 60236.5)$\\
        \hline\hline
    \end{tabular}
\end{table}





\section{Detecting Lensed SNe with ZTF}\label{sec:detecting_glsne}


\subsection{Detection criteria}

\label{sec:detection criteria}


Our simulations provide realistic lightcurves for every supernova observed. To estimate the number of detected supernovae, we need to establish the detection criteria. A single data point is considered a detection only if it passes the 5$\sigma$ threshold, which means that its flux must be at least five times its flux error. In the ZTF, an object is saved as a newly detected source only if it shows at least two 5$\sigma$ detections. However, making follow-up decisions for a SN-like source that shows only 2 detections throughout its entire lightcurve evolution is unrealistic. Therefore, we apply a more stringent criterion on the number of detections, which is at least 5 detections around lightcurve peak (from $-10$ to $+20$ days) that are separated in time more than an hour and spread within at least 5 days. This way we remove transients with significant detections only from high cadence observations in a single night. We set these criteria without any further consideration on the observed bandpasses. We will refer to these as the \textit{detectable} criteria.

However, \textit{detectable} criteria for ZTF are not enough to select a potential lensed candidate. We need to apply selection criterium that identifies potential glSN candidates. For that we need to consider some lensing features. 
In the case of ZTF, we do not expect to resolve the image multiplicity, and therefore we need to focus on the magnification method for identification. 
The magnification method requires that the unresolved lensed SN appears brighter than a normal Type Ia SN at the redshift of the lens galaxy. 
We require that the inferred absolute magnitude, assuming the SN is at the lens redshift, should be brighter than $-19.4$ mag, which is the typical peak magnitude for unlensed Type Ia supernovae. We add this magnitude cut to consider the \textit{identifiable} criterion: 
\begin{equation}
     M_B(t_{\rm peak}) = m_X(t_{\rm peak}) - \mu_D(z_{\rm lens}) - K_{BX} (z_{\rm lens}, t_{\rm peak})<-19.5,
\end{equation} $ m_X(t_{\rm peak})$ correspond to the apparent magnitude in the observed filter $X$ at peak,  $\mu_D(z_{\rm lens})$ is the distance modulus for the lens redshift and $K_{BX}$ the K-correction.

However, there is a different approach to the identification steps depending on a magnitude cut established by the Bright Transient Survey \citep[BTS;][]{2020ApJ...895...32F, Perley2020ApJ...904...35P} to obtain a classification spectrum. BTS is a survey within ZTF aiming to spectroscopically classify all transients up to a magnitude limit, it takes automatic spectra to every transient brigther than 18.5 mag at peak, limit extended to 19 mag but with less completness, in this work we consider the 19 mag the magnitude cut at which we expect to obtain a classification spectrum with ZTF as it matches the case of magnitude range of SN Zwicky.  With a spectrum of the transient, we can determine the redshift, the supernova type, and the phase. That way, we can assess if a supernova is too bright for its redshift and suspect lensing, as was the case for SN Zwicky. On the other hand, if the apparent magnitude does not meet this cut, we need to rely on photometric data and archival redshift measurements either from spectroscopic surveys or photometric redshifts of the lens galaxy. The latter are affected by significant uncertainties that cause bias in the inferred magnitude, as discussed in Sect.~\ref{sec:M_bias_phzerr}. Without the redshift from the SN spectrum, magnified supernovae might look like normal supernovae in ZTF. Therefore, investigations into signatures of lensing and how they impact the lighcurve properties are necessary to identify fainter glSNe.

For this reason, we divide our sample as brighter and fainter than 19 mag at peak; we refer to them as the \textit{bright} and \textit{faint} samples, respectively.


\subsection{Detectability and selection cuts}

Next, we investigate the impact of different selection criteria on the detectability of lensed supernovae. The findings help us develop an optimal search strategy, which we report in Sect.~\ref{sec:conclusion}.

 \subsubsection{Transient magnitude cuts}

\begin{figure}
    \centering
    \includegraphics[width=\linewidth]{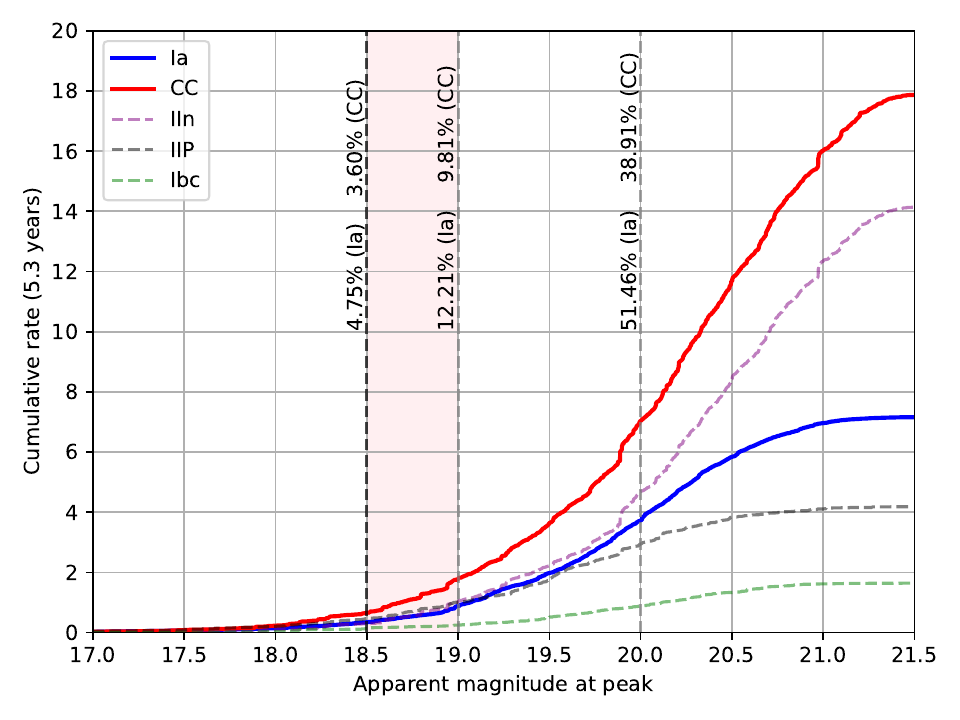}   
    \caption{Expected discovery rate as a function of the apparent magnitude cut for Ia (blue) and CC (red) lensed supernovae. The dashed fainter curves are for subtypes of that are individual components of the red curve for CC; we have considered IIP, IIn, and Ibc types, of which IIn is the dominant type in the lensed CC SN rate from our work. The vertical dashed lines indicate the magnitude cuts of 18.5, 19, and 20 mag with the corresponding percentage of each SN type up to that cut. 18.5 mag corresponds to the BTS magnitude completeness cut, under which most supernovae are spectroscopically classified. We also include a region up to 19 mag because BTS also samples up to 19 mag in some cases, so it would still be possible to get an automatically triggered spectrum.}
    \label{fig:expected_rates}
\end{figure}

We observe a significant relation between the detection rates and the apparent magnitude cut, as can be seen in Fig.~\ref{fig:expected_rates}. Considering the identifiable criteria, we find that only $\sim12$\%  of the lensed Ia supernovae and $\sim10$\% of the CCSNe  are brighter than 19 mag (BTS magnitude cut) corresponding to 0.17 Ia (0.33 CCSN). The curve of the expected discovery rate increases steeply with  magnitude cut, it rapidly increases toward fainter magnitudes. If we set a cut at 20 mag, $\sim52$\% of the SNe Ia are included ($\sim39$\% of the CCSNe), which yields a detection rate of about $0.7$ glSN Ia per year ($1.33$ CCSNe). Over the 5.33 years we thus find $0.88$ Ia ($1.77$ CC) lensed SNe in the Bright Transient Survey, which is consistent with our single discovery of SN Zwicky.

We have examined the sample under the detectable criteria in order to determine how many glSNe we might be missing due to not passing the absolute brightness threshold. Our brightness cut-off is set at an intrinsic brightness level higher than that of a normal SNe Ia ($-19.5$ mag) to identify lensed SNe. From our analysis, we found that about 20\% of Ia and 23\% of CC supernovae from the detectable sample can not be identified, because of the absolute magnitude inferred when assuming the lens redshift. This indicates that there may be around 4 Ia (12 CC) glSNe in the ZTF data that cannot be identified using the magnification method in 5.33 years.

\begin{figure*}[h]
    \centering
    \includegraphics[width=\linewidth]{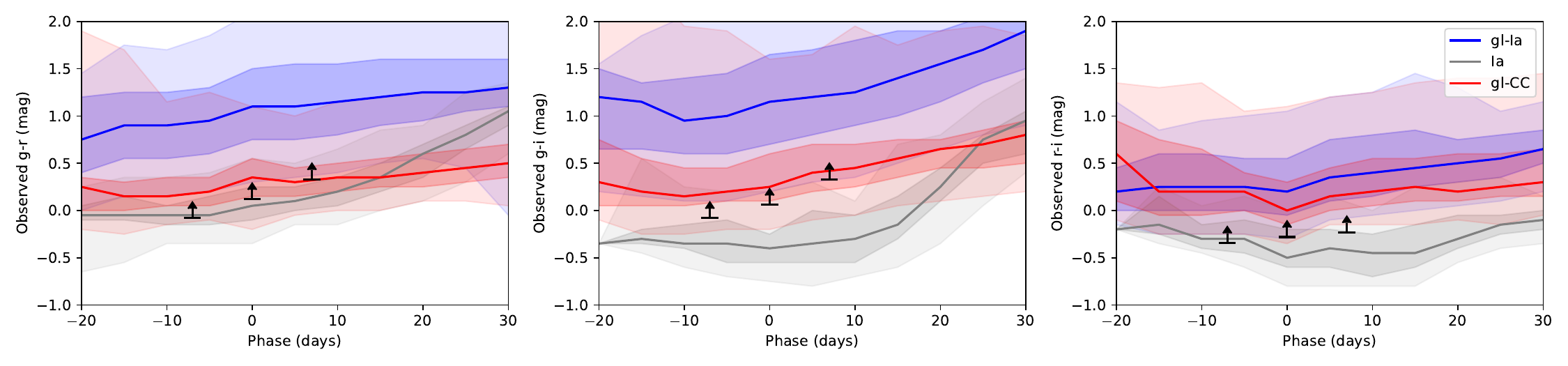}
    \caption{Measured color evolution for the simulated lightcurves. I would reword as "From left to right, we show the observed $g-r$, $g-i$ and $r-i$ colors for lensed SNe Ia (blue), unlensed SN Ia (grey), and lensed CCSNe (red). The black lower limits show the color cut extracted in this study that retain 90\% of the Ia lensed events and exclude the largest possible amount of candidate contaminants from normal SNe Ia.} 
    \label{fig:color}
\end{figure*}

\begin{table}
    \centering
    \caption{Optimal selection cut from glSN lightcurves. $t_0$ denotes time of maximum light.  $g-i$, $g-r$, $r-i$ observed colours are corrected from the MW extinction. SALT2 fits include all the available dertections for each SN. }
    \begin{tabular}{cccc}
         \hline
         \hline 
        \\
        \multicolumn{4}{p{.6\linewidth}}{Observed transient colour cuts} 
        \\
         \hline
            Phase (days)  &  $g-i$ & $g-r$ & $r-i$ \\
            $t_0 - 7$ & $>-0.08$ & $>-0.08$ & $>-0.34$ \\
            $t_0$ & $>0.06$ & $>0.12$ & $>-0.28$ \\
            $t_0 + 7$ & $>0.33$ & $>0.33$ & $>-0.23$ \\
        \hline
        \\
        \multicolumn{4}{p{.6\linewidth}}{SALT2 fits} 
        \\
         \hline
         Colour parameter    & $c>0$ & &  \\
         Absolute Magnitude & $M_B<-20$ & &  \\
         \hline
         \hline
    \end{tabular}
    \label{tab:cuts}
\end{table}

\subsubsection{Observed transient colour}

Supernovae that are gravitationally lensed occur at a higher redshift than those that are not lensed. As a result, their observed colour is expected to be redder than that of unlensed supernovae at the {\em lens} redshift. Furthermore, due to time dilation, their colour evolution appears to be slower in the observer frame.

We compare the synthetic lensed lightcurves to a simulated sample of unlensed SNe Ia and investigate optimal selection cuts. We measure the observed colour at maximum and one week before and after the peak using three ZTF filter combinations: $g-r$, $g-i$, and $r-i$. The synthetic Type Ia and CC lensed SNe and the typical unlensed Type Ia SN are shown in Fig.~\ref{fig:color} with the $g-r$, $g-i$, and $r-i$ per observed epoch.  

\begin{figure}
    \centering
    \includegraphics[width=\linewidth]{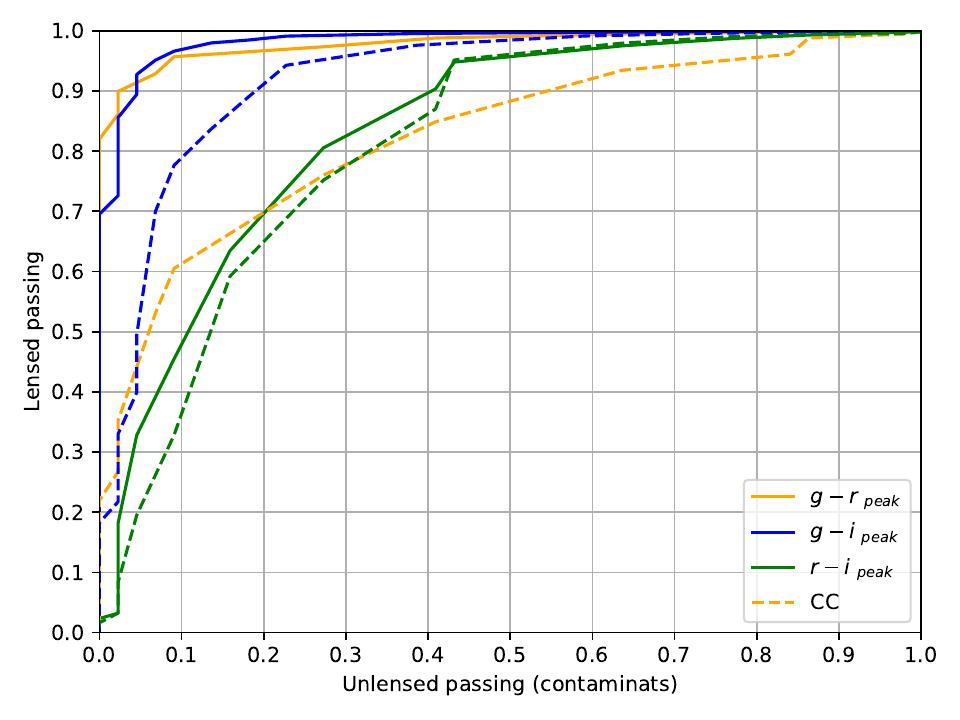}
    \caption{Percentage of glSN passing the colour cuts versus the percentage of normal SNe Ia that is kept. Different colours indicate the colour combination: $g-r$ in yellow, $g-i$ in blue, and $r-i$ in green. Solid lines indicates the results of distinguishing normal SNe Ia from lensed SNe Ia, while the dashed lines compare normal SNe Ia with lensed CCSNe.}
    \label{fig:true_false_color}
\end{figure}

To investigate optimal selection cuts in colour space we look for cuts that better differentiate lensed from unlensed SNe. This investigation is targeted for SNe Ia, but CCSN are also included to understand how many would pass our cuts. We consider unlensed Ia contaminants and glSNe true candidates, and we study the percentage of events passing colour cuts at the peak and one week before and after lightcurve maximum. The resulting true versus false candidate percentages per colour cut is visualized in Fig.~\ref{fig:true_false_color}. It is noticeable that the most efficient colour distinction is $g-i$, but the other two colours also provide promising results.

We examine cuts that retain 90\% of the lensed Ia and look at how many lensed CCSN and unlensed Ia pass the cut, the latter being considered a contaminant that we want to minimize. The set cuts are summarized in table \ref{tab:cuts}. The best colours cut is the combination of $g-i >0.06$, $g-r > 0.12$ and $r-i>-9.28$ at peak, $g-i > -0.08$, $g-r > -0.08$, $r-i > -0.34$ one week before peak and : $g-i >  0.33$, $g-r > 0.33$, $r-i > -0.23$ one week after peak brightness. We find that 70\%, 56\% and 92\% of lensed CCSN pass the $g-i$, $g-r$ and $r-i$ cuts, respectively, while 10\%, 30\% and 40\% for the unlensed SNe Ia. 
The extracted cuts are illustrated in Fig.~\ref{fig:color} with black arrows. 
These cuts are employed in T24 to select photometric lensed candidates.

However, not all lightcurves present the same colour information. For lensed SNe, $g$ is often undetected, and $i$ is less often observed than $r$ and $g$. Approximately 17\% of the lensed Ia show $g-r$ colour at any of the 3 epochs considered, approximately 9\% $g-i$, and 33\% $r-i$. The percentage of lensed SNe Ia passing any of the colour cuts at any of the 3 epochs considered is approximately 42\%, or one per year. Similarly, for lensed CC, we found that approximately 61\% or 2.3 per year pass the cuts.

\subsubsection{SALT2 parameters}

SNe Ia are often modelled using the SALT2 empirical lightcurve model. By fitting Ia lightcurves with SALT2, we obtain an estimate of the stretch ($x_1$) that describes the broadness of the lightcurve, color ($c$) including intrinsic color of the SN and dust, time of maximum ($t_0$) and the inferred absolute peak magnitude in B-rest frame  ($M_B$). 
The SALT2 model can have converging fits to a variety of shorter duration SN-like transients, so it is a simple metric for evaluating live/archival candidates as it was applied in T24.
We investigate how the SALT2 model performs on glSNe and we explored how selection cuts of the SALT2 fit parameters could identify potential lensed candidates.

\begin{figure}
    \centering
    \includegraphics[height=.588\linewidth]{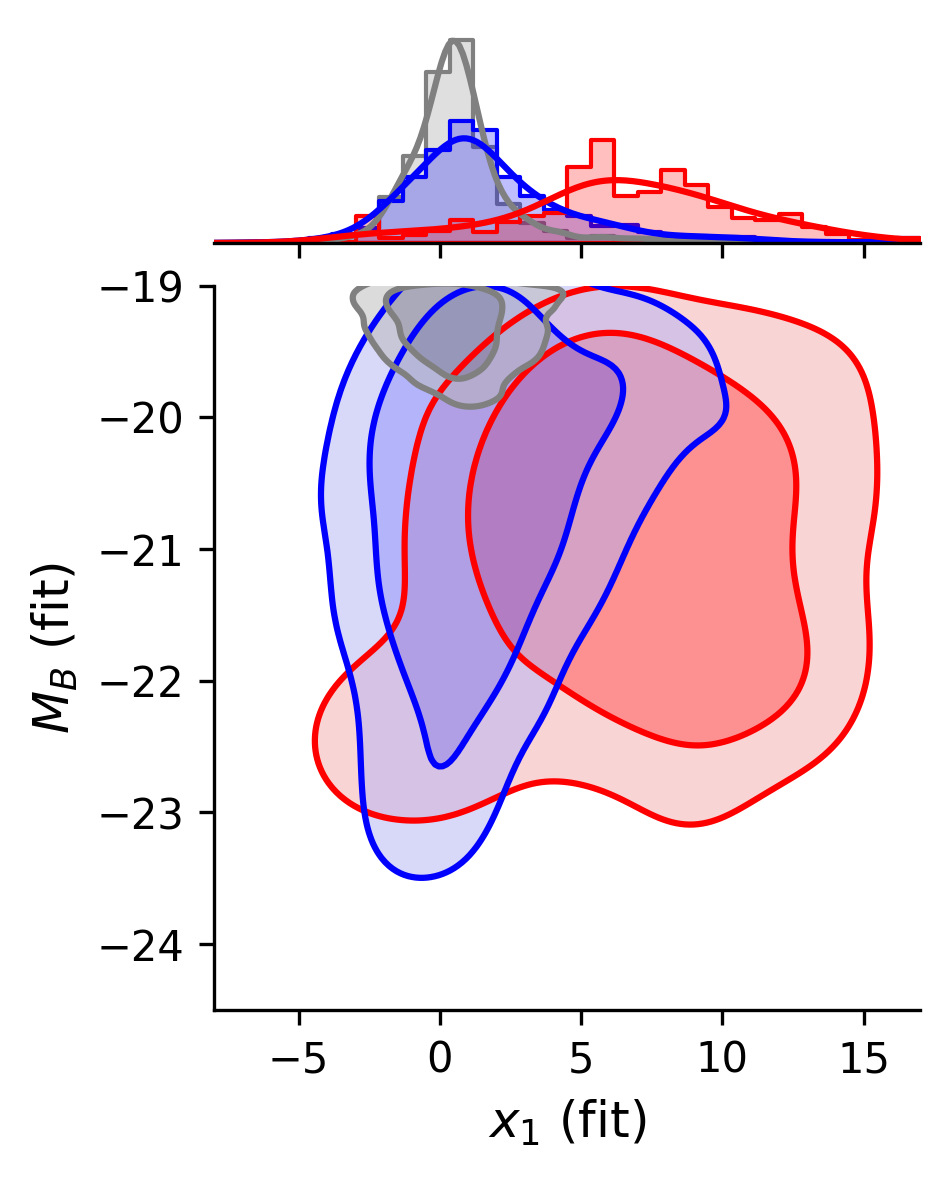}
    \includegraphics[height=.588\linewidth]{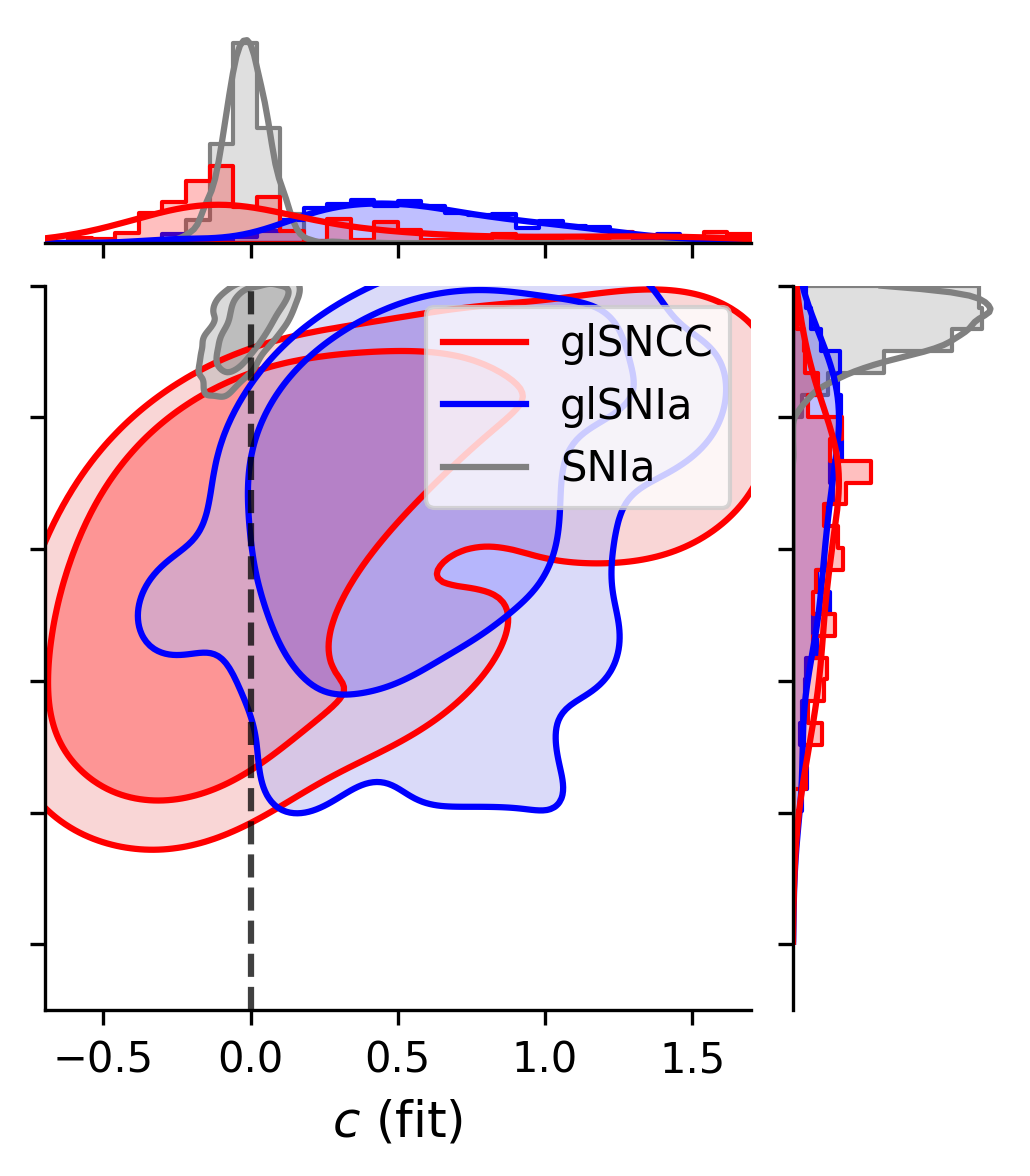}
    \caption{Joint distribution of the best fit parameters from SALT2 model to the identifiable sample. On the yaxis the infered absolute magnitude ($M_B$), and on the xaxis we show the stretch ($x_1$) and the colour parameters ($c$). The contours mark the 1$\sigma$, and the 2$\sigma$ regions. Plotted for unlensed Ia (gray), lensed SNIa (blue) and lensed CCSN (red). 
    }
    \label{fig:salt2_fit}
\end{figure}

Figure \ref{fig:salt2_fit} shows the distribution of the fitted SALT2 parameters for glSN Ia and CC as well as for normal SNe Ia. 
The stretch distribution shows a tail towards higher values, but there is not a clear cut in that parameter with 1.8 mean value for glSNe Ia. On the other hand, fitting SALT2 to glCCSNe generally returns significantly larger $x_1$ values with a mean value of 6.7. In colour space, it is noticeable that the lensed Ia colour distribution is shifted to larger (redder) values. We find that $\sim90$\% of the glSNe Ia show $c>0$, this is another selection cut also implemented in T24, which excludes around half of the unlensed events and most of the glCCSNe. If we combine the parameters with the inferred $M_B$, there is a clear separation, we observe that setting a cut of $M_B<-20$ mag excludes almost all the unlensed SNe Ia, even though losing close to $\sim20$\% of the identifiable sample. 

Examining the number of events passing the cuts from the SALT2 fit parameters we obtain: $\sim90$\% (2.15 per year) pass $c>0$, $\sim80$\% (1.9 per year) pass $M_B<-20$ mag.

\begin{figure*}
    \centering
    \includegraphics[height=5.88cm]{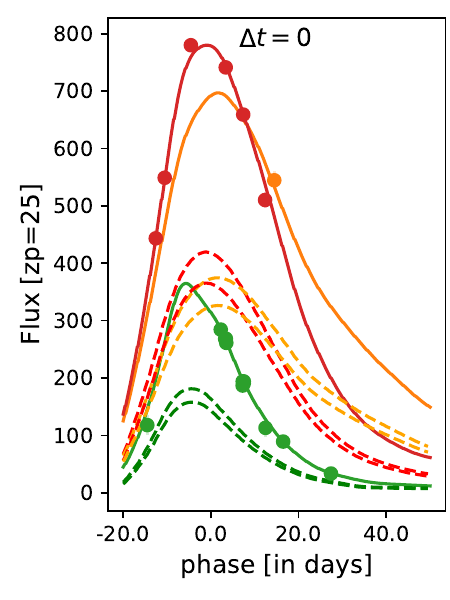}
    \includegraphics[height=5.88cm]{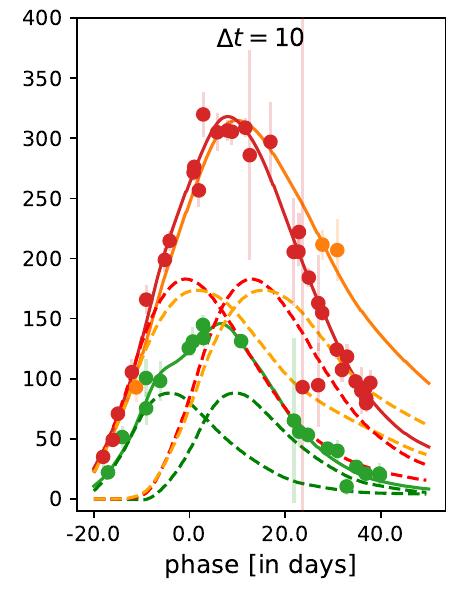}
    \includegraphics[height=5.88cm]{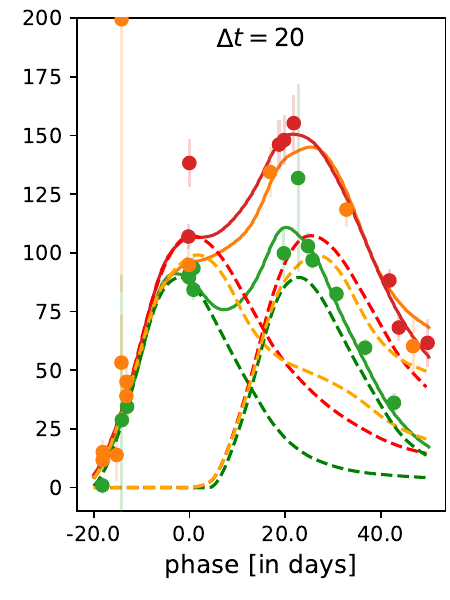}
    \includegraphics[height=5.88cm]{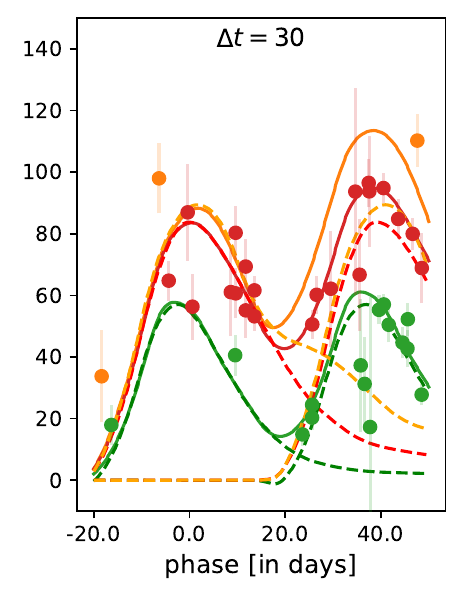}
    \caption{Unresolved lightcurves in $g$ (green), $r$ (red) and $i$- bands (orange) for a simulated doubly-imaged SNe Ia with $x_1=0$ and $c=0$. To illustrate the effect on the unresolved lightcurve of the added magnification and the time delay between individual images, this plot shows four examples with similar flux ratios and time delays of 0, 10, 20 and 30 days.}
    \label{fig:lcs_compare}
\end{figure*}
\section{The impact of time delays}\label{sec:analysis method}

Observations of unresolved glSN lightcurves correspond to the superposition of multiple images, where each individual image has a distinct magnification and arrival time. The combined lightcurve appears brighter than in the resolved case because of the addition of flux from multiple lightcurves. The shape depends on the time delay, which can broaden the lightcurve or even create multiple bumps in the case of significant time delays.  Figure~\ref{fig:lcs_compare} provides an illustration of the effect of time delays for a SN Ia.

The average values for lensing parameters including source and lens redshift, total magnification, time delays and Einstein radius are presented in Tab.~\ref{tab:det_distr_stat} for the detected simulated lightcurves. We distinguish between the faint and bright sample in this section.

\begin{table*}
    \centering
    \caption{Statistical results on the detection rates and the lensing parameters of the simulated SN in this work. We distinguish the sample by the detection criteria and the faint and bright as defined in Sect.~\ref{sec:detection criteria}. The uncertainties correspond to the 1$\sigma$ confidence level. }
    \label{tab:det_distr_stat}
    \begin{tabular}{l|ccccccc|}
        \hline\hline
        & \#SN (5.33 yrs) & \#SN ($\rm yr^{-1}$) & $\rm z_{source}$ & $\rm z_{lens}$ & $\rm \Delta mag$ (mag) & $\rm \Delta t_{max}$ (days) & $\rm \theta_{Ein}$ ('') \\
        \hline
        \textit{Detectable} sample & & & & & & & \\
        glSN-Ia (all)     & 21.05 & 3.95 & $ 0.60 ^{+ 0.26 }_{ -0.24 }$ & $ 0.21 ^{+ 0.13 }_{ -0.14 }$ & $ 2.94 ^{+ 1.29 }_{ -1.24 }$ & $ 6.83 ^{+ 16.87 }_{ -4.70 }$ & $ 0.76 ^{+ 0.44 }_{ -0.41 }$ \\
        glSN-Ia ($m<19$)  & 1.39 & 0.26 & $ 0.56 ^{+ 0.31 }_{ -0.29 }$ & $ 0.18 ^{+ 0.12 }_{ -0.12 }$ & $ 4.37 ^{+ 1.12 }_{ -1.20 }$ & $ 5.24 ^{+ 14.98 }_{ -3.92 }$ & $ 0.80 ^{+ 0.47 }_{ -0.41 }$ \\
        glSN-Ia ($m>19$)  & 19.67 & 3.69 & $ 0.61 ^{+ 0.25 }_{ -0.24 }$ & $ 0.20 ^{+ 0.13 }_{ -0.12 }$ & $ 2.85 ^{+ 1.24 }_{ -1.28 }$ & $ 6.92 ^{+ 21.24 }_{ -4.62 }$ & $ 0.76 ^{+ 0.44 }_{ -0.41 }$ \\
        glSN-CC (all)     & 56.02 & 10.51 & $ 0.76 ^{+ 0.41 }_{ -0.37 }$ & $ 0.24 ^{+ 0.15 }_{ -0.14 }$ & $ 3.03 ^{+ 1.20 }_{ -1.35 }$ & $ 8.64 ^{+ 27.45 }_{ -5.82 }$ & $ 0.82 ^{+ 0.48 }_{ -0.47 }$ \\
        glSN-CC ($m<19$)  & 2.88 & 0.54 & $ 0.66 ^{+ 0.49 }_{ -0.36 }$ & $ 0.18 ^{+ 0.13 }_{ -0.11 }$ & $ 4.46 ^{+ 1.61 }_{ -1.79 }$ & $ 5.04 ^{+ 15.46 }_{ -3.29 }$ & $ 0.87 ^{+ 0.41 }_{ -0.50 }$ \\
        glSN-CC ($m>19$)  & 53.14 & 9.97 & $ 0.77 ^{+ 0.41 }_{ -0.35 }$ & $ 0.24 ^{+ 0.17 }_{ -0.15 }$ & $ 2.98 ^{+ 1.23 }_{ -1.21 }$ & $ 8.86 ^{+ 19.82 }_{ -6.20 }$ & $ 0.81 ^{+ 0.49 }_{ -0.46 }$ \\
        \hline
        \textit{Identifiable} sample & & & & & & & \\
        glSN-Ia (all)                & 7.25 & 1.36 & $ 0.68 ^{+ 0.23 }_{ -0.23 }$ & $ 0.34 ^{+ 0.13 }_{ -0.12 }$ & $ 3.60 ^{+ 0.99 }_{ -1.15 }$ & $ 8.71 ^{+ 20.91 }_{ -5.90 }$ & $ 0.58 ^{+ 0.37 }_{ -0.34 }$
 \\
        glSN-Ia ($m<19$)             & 0.91 & 0.17 & $ 0.58 ^{+ 0.31 }_{ -0.25 }$ & $ 0.24 ^{+ 0.14 }_{ -0.11 }$ & $ 4.56 ^{+ 0.93 }_{ -1.04 }$ & $ 6.92 ^{+ 30.67 }_{ -5.06 }$ & $ 0.70 ^{+ 0.40 }_{ -0.37 }$ \\
        glSN-Ia ($m>19$)             & 6.34 & 1.19 & $ 0.69 ^{+ 0.22 }_{ -0.22 }$ & $ 0.35 ^{+ 0.12 }_{ -0.11 }$ & $ 3.45 ^{+ 1.02 }_{ -1.00 }$ & $ 8.79 ^{+ 25.71 }_{ -6.52 }$ & $ 0.56 ^{+ 0.33 }_{ -0.33 }$ \\
        glSN-CC (all)                & 16.42 & 3.08 & $ 0.93 ^{+ 0.38 }_{ -0.37 }$ & $ 0.41 ^{+ 0.17 }_{ -0.17 }$ & $ 3.74 ^{+ 1.05 }_{ -1.07 }$ & $ 9.98 ^{+ 33.13 }_{ -7.56 }$ & $ 0.61 ^{+ 0.33 }_{ -0.36 }$ \\
        glSN-CC ($m<19$)             & 1.71 & 0.32 & $ 0.72 ^{+ 0.44 }_{ -0.36 }$ & $ 0.24 ^{+ 0.12 }_{ -0.11 }$ & $ 4.74 ^{+ 1.47 }_{ -1.65 }$ & $ 5.40 ^{+ 19.43 }_{ -3.23 }$ & $ 0.79 ^{+ 0.47 }_{ -0.44 }$ \\
        glSN-CC ($m>19$)             & 14.71 & 2.76 & $ 0.96 ^{+ 0.39 }_{ -0.37 }$ & $ 0.42 ^{+ 0.16 }_{ -0.15 }$ & $ 3.63 ^{+ 0.99 }_{ -1.04 }$ & $ 11.04 ^{+ 37.74 }_{ -8.41 }$ & $ 0.59 ^{+ 0.35 }_{ -0.33 }$ \\
        
        \hline
        \hline\hline
    \end{tabular}
\end{table*}

\subsection{Source and lens redshifts}

The average source redshift for the identifiable sample is $z=0.61$ for Ia and $z=0.77$ for CC. The average redshift of the galaxy acting as a lens is 0.34 for Ia and 0.41 for CCSNe. These are much higher redshifts than for unlensed supernovae in ZTF, which are typically found at redshifts $z<0.1$. Therefore, it makes sense to exclude candidates with small redshifts $z<0.1$ to avoid unlensed supernova contaminants, given the limited follow-up resources, which cut is also applied in T24. 

\begin{figure}
    \centering
    \includegraphics[width=\linewidth]{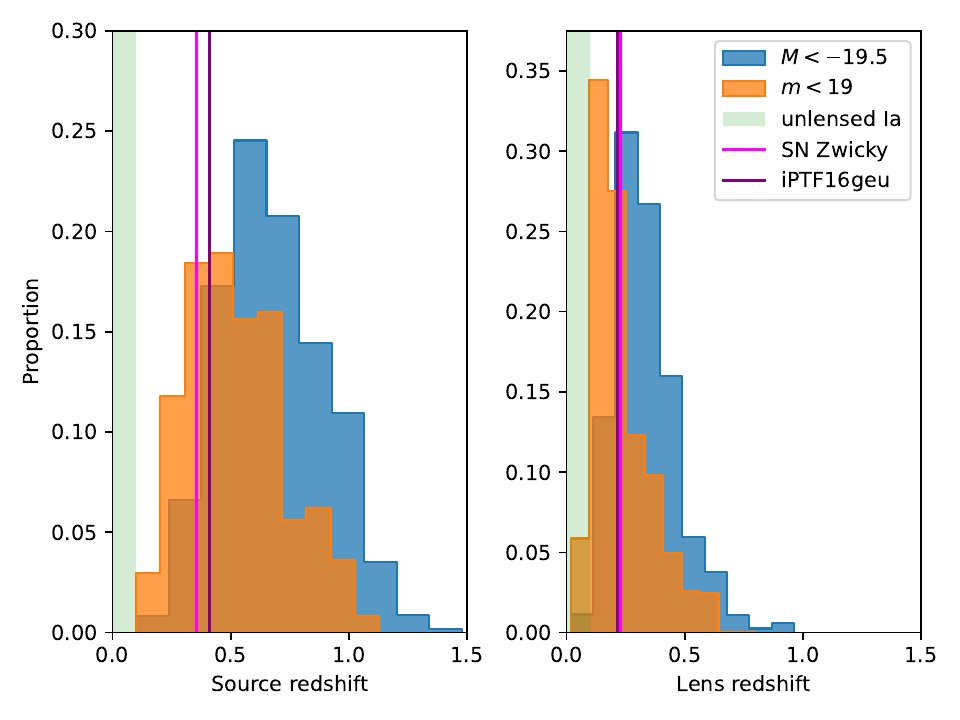}
    \caption{The distribution of glSN-Ia source redshifts (left) and lens redshifts (right) for glSNe Ia that pass the identifiable criteria (blue) and glSNe in the bright sample with $m<19$ (orange). The green area illustrates the redshift range ($z<0.1$) for unlensed Ia supernovae detected in ZTF. The vertical lines show the values for SN Zwicky (pink) and iPTF16geu (purple).}
    \label{fig:zs_zl}
\end{figure}

The only identified lensed Ia within ZTF is SN Zwicky with $z_{source} = 0.3544$ and $z_{lens}=0.2262$. These redshifts are comparable to the average for the bright sample that would be detected in a similar way as SN Zwicky, given that the brightness cut ensures classification spectra from the BTS  survey (see Fig. \ref{fig:zs_zl}). 

\begin{figure}
    \centering
    \includegraphics[width=\linewidth]{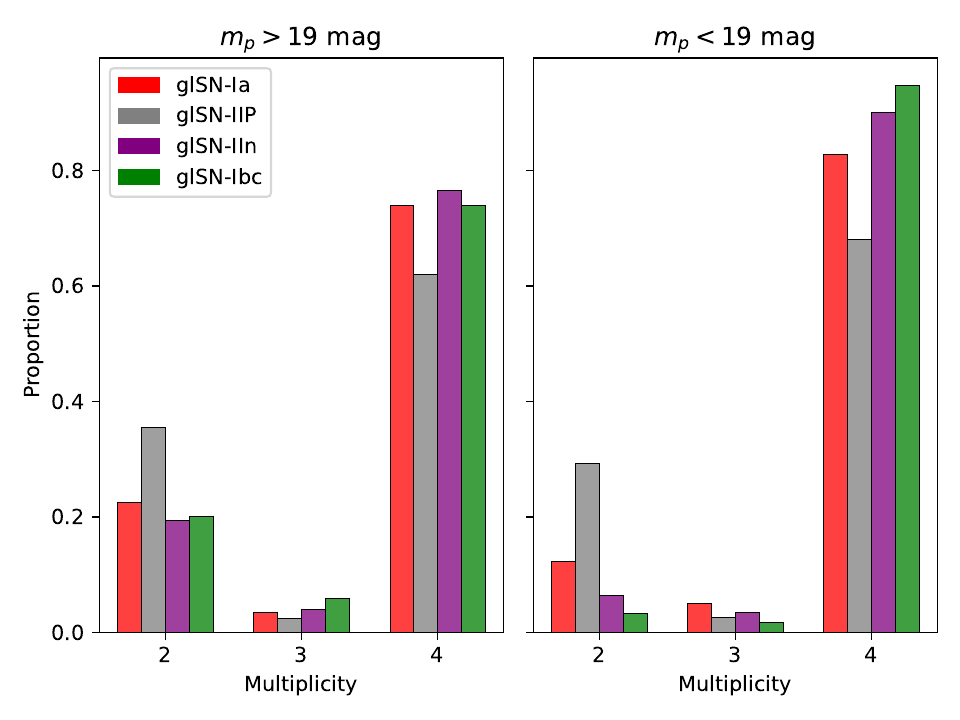}
    \caption{The distribution of multiplicity for the identifiable sample, shown for the faint (left) and bright (right) samples. We plot for lensed Ia and the subtypes of CCSN simulated in this work.}
    \label{fig:multiplicity}
\end{figure}

\subsubsection{Image multiplicity}\label{subsec:multiplicity}

The simulated lensing systems show three types of multiplicity with 2, 3 and 4 images, that we refer to as doubles, triplets and quads, respectively.

The lens simulations made with \texttt{lenstronomy} provided a relative proportion of $\sim$87\% doubles, $\sim$1\% triplets and $\sim$12\% quads, implying that doubles are more commonly produced in gravitational lensing systems. Nevertheless, higher multiplicity generally translates to higher total magnifications. In the case of unresolved lightcurves, the observed magnitude corresponds to the sum of the individual fluxes. In that case higher total magnifications provide apparent brighter lightcurves that are also more likely to be detected for a ground-based survey like ZTF. 
Figure \ref{fig:multiplicity} illustrates the relative fraction of multiplicity per supernova type for the faint and the bright subsample of the identifiable glSNe. Four images are the most likely scenario in the faint and bright regimes, and across the SN types, in agreement with \cite{Goldstein2019ApJS..243....6G}.
75\% of the identifiable glSNe Ia are quads, 21\% doubles and 3\% triples, considering only the bright sample with peak magnitude less than 19 mag, the relative fraction becomes 85\% quads, 10\% doubles and 5\% triplets, indicating the preference of quads with brighter magnitude cuts.

\subsubsection{System properties: Time delay, magnification and image separation}

Next, we examine the distributions of the lensing parameters for the identifiable sample of Type Ia glSNe.  The injected distribution of magnifications is an exponentially decaying function favoring low values (see Fig.~\ref{fig:input_lensparams}). The average magnification from our sample is 3.59 mag, corresponding to $\mu_{total} \sim 27.29$ \footnote{The minimum magnification required for strong lensing is 2.}.  For the bright sample, it becomes 4.55 mag or $\mu_{total}\sim 66$. This shows a selection bias towards highly magnified events in ZTF. This result agrees with the findings by  \citet{Sainz2023MNRAS.526.4296S}.  

The maximum time delay between the multiple images correlates inversely with the magnification. The magnification is larger for events with more symmetric geometrical alignment between  source and lens. These cases will have shorter differences in arrival times. Hence, the bias towards highly magnified events also favours short time delays. In our sample the average maximum time delay is 8.5 days, 7.1 days for magnitude limit $m<19$ mag. 

SN Zwicky and iPTF16geu were both highly magnified, with small time delays around zero days. SN Zwicky was magnified $\Delta m =  3.44\pm0.14$ and iPTF16geu $\Delta m = 4.3 \pm 0.2$. The values are within the expectations for 50\% of our simulation for the bright sample.

\section{Discussion and conclusions}\label{sec:conclusion}

In this work, we simulated lensed Type Ia supernovae and core-collapse supernovae. We assume a recipe of lens galaxies as detailed in Sect.~\ref{sec:lensing_sim} and combine this with volumetric supernova rates and the probability of lensing per source redshift to obtain a distribution of lensed supernovae. From the modelled systems, we simulated realistic lightcurves by combining with the ZTF observing logs over 5.33 years of the survey. 

We distinguished 4 categories of detections in this work, differentiated by the detection criteria: detectable and identifiable; and by the apparent magnitude cut from BTS: bright and faint. First, in regard to the lightcurve requirements, we distinguish the detectable sample as the events that pass 5 detections around peak, and the identifiable sample as those that also satisfy the magnitude criteria for the magnification method, meaning that their inferred absolute magnitude is brighter than normal SNe Ia. 

The predicted rates for ZTF are revised in this work and presented in Table~\ref{tab:det_distr_stat}. We find that, in the identifiable sample, 1.36 glSNe Ia and 3.08 glSNe CC are predicted per year, from which 0.17 glSNe Ia will be in the bright sample and 1.19 in the faint sample. For glSNe CC, the values are 0.32 and 2.76. In the 5.33 years of the survey, around 1 glSN are brighter than 19 mag, consistent with finding only SN Zwicky. The finding of 1 glSN Ia brighter than 19 mag is thus consistent with our simulations. T24 conduct an archival search for lensed supernova in ZTF, with candidates landing mostly in the magnitude range fainter than the BTS cut. Based on the simulations presented in this work, 3.87 Ia and 8.97 CC glSNe should occur in the faint sample from which 2.17 glSNeIa and 0.6 glCCSNe pass the cuts summarize in Tab, \ref{tab:cuts} during 3.25 years which corresponds to the search time span in T24.

If we look at the magnification versus source redshift, SN Zwicky and iPTF16geu are somehow outliers if compared with the whole distribution, but they are highly probable and expected in the bright sample. This indicates a bias in ZTF for classifying highly magnified events because of the apparent magnitude cut from the BTS survey.

\begin{figure*}
    \centering
    \includegraphics[height=.35\linewidth]{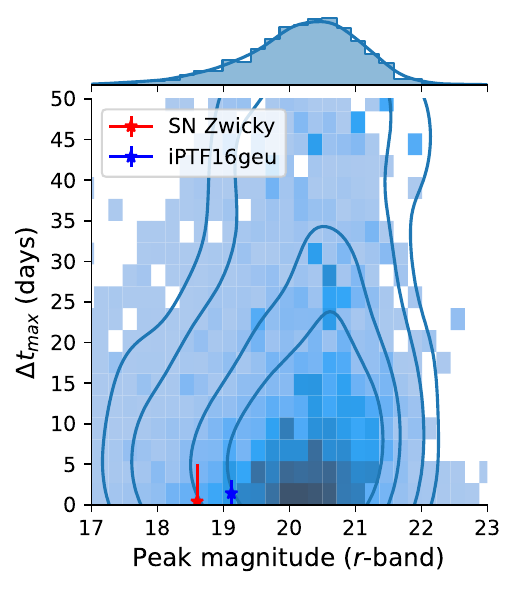}
    \includegraphics[height=.35\linewidth]{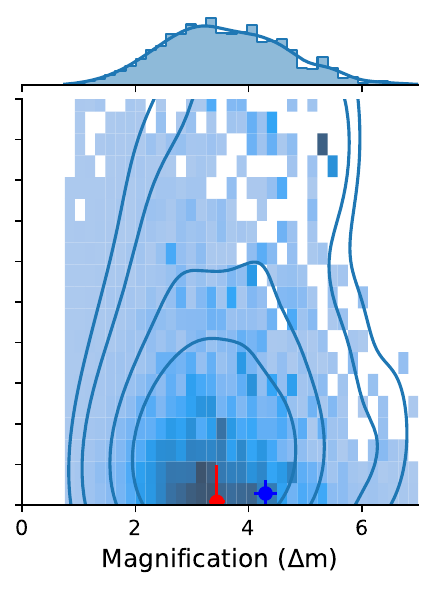}
    \includegraphics[height=.35\linewidth]{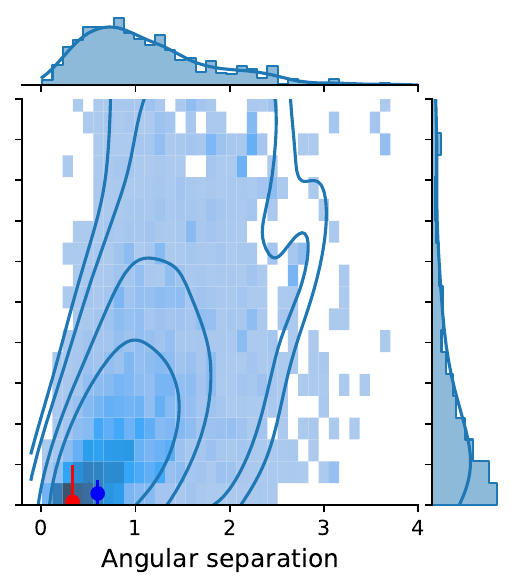}
    \caption{These panels show the relations of time delay in days with the peak magnitude, total magnification and angular separations (in arcsecond). The stars in red and purple correspond to the values measured from SN Zwicky and iPTF16geu, respectively. The contours show the 50\% 68\% 90\% and 95\% percentiles. }
    \label{fig:r_td_max}
\end{figure*}

\begin{figure}
    \centering
    \includegraphics[width=\linewidth]{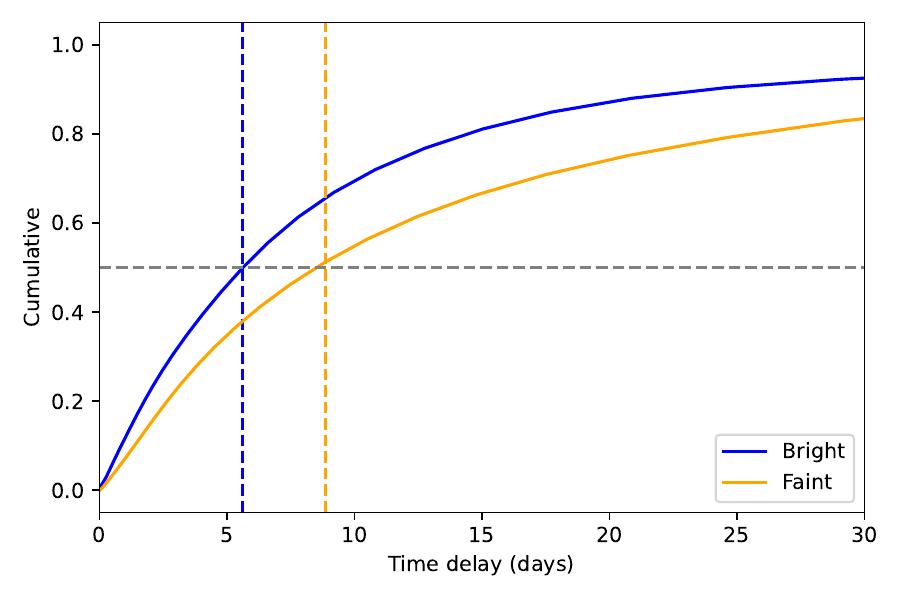}
    \caption{Cumulative distribution normalized to 1 of the time-delay for the Faint and the bright sample. The dashed lines show the median for both samples that correspond to 5.6 days for the bright sample and 8.9 for the Faint sample.}
    \label{fig:td_cum}
\end{figure}

Even though we see in Fig.~\ref{fig:r_td_max} that the mode of the glSN time delay distribution is below 5 days, the cumulative distribution  in Fig.~\ref{fig:td_cum} indicates that more than 50\% of the faint show more than 8.9 days and 50\% of the Bright sample more than 5.6 days, from which 10\% or better  measurements of time delays are feasible. 

\begin{figure}
    \centering
    \includegraphics[width=.9\linewidth]{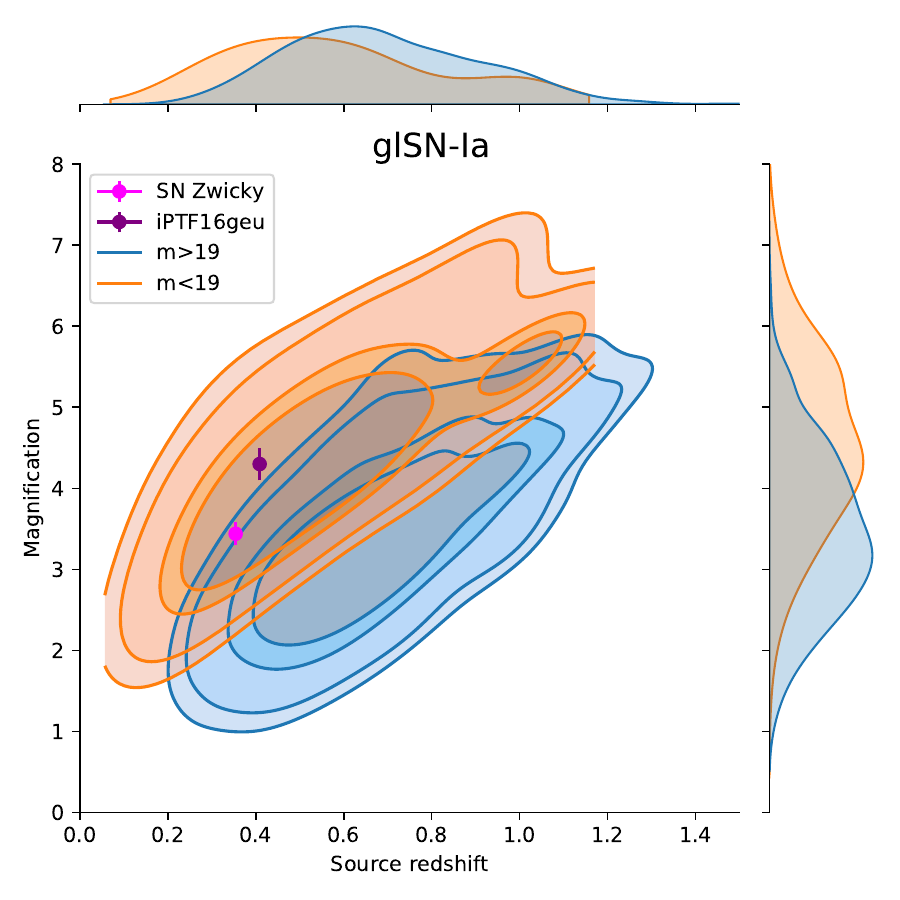}
    \caption{The distribution of source redshift and magnification in magnitudes for glSN-Ia. 
    The orange contour shows events with lightcurves peaking brighter than 19 mag while the blue contour shows the rest of the sample. SN Zwicky and iPTF16geu are overplotted. They appear outliers when considering the faint sample that corresponds to a larger detection rate. However considering the magnitude bias from the BTS magnitude cut, they fall within the 50\% contour. 
    }
    \label{fig:dm_zs}
\end{figure}

\begin{acknowledgements}
        This work has been supported by the research project grant “Understanding the Dynamic Universe” funded by the Knut and Alice Wallenberg Foundation under Dnr KAW 2018.0067,  {\em Vetenskapsr\aa det}, the Swedish Research Council, project 2020-03444. and the G.R.E.A.T research environment, project number 2016-06012.
        S. Schulze is partially supported by LBNL Subcontract NO. 7707915.
\end{acknowledgements}

%
%

\bibliographystyle{aa}
\bibliography{aanda.bib}

\begin{thebibliography}{55}
\expandafter\ifx\csname natexlab\endcsname\relax\def\natexlab#1{#1}\fi

\bibitem[{{Arendse} {et~al.}(2023){Arendse}, {Dhawan}, {Sagu{\'e}s Carracedo},
  {Peiris}, {Goobar}, {Wojtak}, {Alves}, {Biswas}, {Huber}, {Birrer}, \& {The
  LSST Dark Energy Science Collaboration}}]{Arendse2023arXiv231204621A}
{Arendse}, N., {Dhawan}, S., {Sagu{\'e}s Carracedo}, A., {et~al.} 2023, arXiv
  e-prints, arXiv:2312.04621

\bibitem[{{Barbary} {et~al.}(2016){Barbary}, {Barclay}, {Biswas}, {Craig},
  {Feindt}, {Friesen}, {Goldstein}, {Jha}, {Rodney}, {Sofiatti}, {Thomas}, \&
  {Wood-Vasey}}]{Barbary2016ascl.soft11017B}
{Barbary}, K., {Barclay}, T., {Biswas}, R., {et~al.} 2016, {SNCosmo: Python
  library for supernova cosmology}, Astrophysics Source Code Library, record
  ascl:1611.017

\bibitem[{{Bellm} {et~al.}(2019){Bellm}, {Kulkarni}, {Graham}, {Dekany},
  {Smith}, {Riddle}, {Masci}, {Helou}, {Prince}, {Adams}, {Barbarino},
  {Barlow}, {Bauer}, {Beck}, {Belicki}, {Biswas}, {Blagorodnova}, {Bodewits},
  {Bolin}, {Brinnel}, {Brooke}, {Bue}, {Bulla}, {Burruss}, {Cenko}, {Chang},
  {Connolly}, {Coughlin}, {Cromer}, {Cunningham}, {De}, {Delacroix}, {Desai},
  {Duev}, {Eadie}, {Farnham}, {Feeney}, {Feindt}, {Flynn}, {Franckowiak},
  {Frederick}, {Fremling}, {Gal-Yam}, {Gezari}, {Giomi}, {Goldstein},
  {Golkhou}, {Goobar}, {Groom}, {Hacopians}, {Hale}, {Henning}, {Ho}, {Hover},
  {Howell}, {Hung}, {Huppenkothen}, {Imel}, {Ip}, {Ivezi{\'c}}, {Jackson},
  {Jones}, {Juric}, {Kasliwal}, {Kaspi}, {Kaye}, {Kelley}, {Kowalski},
  {Kramer}, {Kupfer}, {Landry}, {Laher}, {Lee}, {Lin}, {Lin}, {Lunnan},
  {Giomi}, {Mahabal}, {Mao}, {Miller}, {Monkewitz}, {Murphy}, {Ngeow},
  {Nordin}, {Nugent}, {Ofek}, {Patterson}, {Penprase}, {Porter}, {Rauch},
  {Rebbapragada}, {Reiley}, {Rigault}, {Rodriguez}, {van Roestel}, {Rusholme},
  {van Santen}, {Schulze}, {Shupe}, {Singer}, {Soumagnac}, {Stein}, {Surace},
  {Sollerman}, {Szkody}, {Taddia}, {Terek}, {Van Sistine}, {van Velzen},
  {Vestrand}, {Walters}, {Ward}, {Ye}, {Yu}, {Yan}, \&
  {Zolkower}}]{Bellm2019PASP..131a8002B}
{Bellm}, E.~C., {Kulkarni}, S.~R., {Graham}, M.~J., {et~al.} 2019, \pasp, 131,
  018002

\bibitem[{{Birrer} \& {Amara}(2018)}]{Birrer2018PDU....22..189B}
{Birrer}, S. \& {Amara}, A. 2018, Physics of the Dark Universe, 22, 189

\bibitem[{{Birrer} {et~al.}(2021){Birrer}, {Shajib}, {Gilman}, {Galan},
  {Aalbers}, {Millon}, {Morgan}, {Pagano}, {Park}, {Teodori}, {Tessore},
  {Ueland}, {Van de Vyvere}, {Wagner-Carena}, {Wempe}, {Yang}, {Ding},
  {Schmidt}, {Sluse}, {Zhang}, \& {Amara}}]{Birrer2021JOSS....6.3283B}
{Birrer}, S., {Shajib}, A., {Gilman}, D., {et~al.} 2021, The Journal of Open
  Source Software, 6, 3283

\bibitem[{{Cardelli} {et~al.}(1989){Cardelli}, {Clayton}, \&
  {Mathis}}]{Cardelli1989}
{Cardelli}, J.~A., {Clayton}, G.~C., \& {Mathis}, J.~S. 1989, \apj, 345, 245

\bibitem[{{Chen} {et~al.}(2022){Chen}, {Kelly}, {Oguri}, {Broadhurst}, {Diego},
  {Emami}, {Filippenko}, {Treu}, \& {Zitrin}}]{Chen2022Natur.611..256C}
{Chen}, W., {Kelly}, P.~L., {Oguri}, M., {et~al.} 2022, \nat, 611, 256

\bibitem[{{Choi} {et~al.}(2007){Choi}, {Park}, \&
  {Vogeley}}]{Choi2007ApJ...658..884C}
{Choi}, Y.-Y., {Park}, C., \& {Vogeley}, M.~S. 2007, \apj, 658, 884

\bibitem[{{Collett}(2015)}]{Collett2015ApJ...811...20C}
{Collett}, T.~E. 2015, \apj, 811, 20

\bibitem[{{Dark Energy Survey Collaboration}(2016)}]{des}
{Dark Energy Survey Collaboration}. 2016, \mnras, 460, 1270

\bibitem[{{Dilday} {et~al.}(2008){Dilday}, {Kessler}, {Frieman}, {Holtzman},
  {Marriner}, {Miknaitis}, {Nichol}, {Romani}, {Sako}, {Bassett}, {Becker},
  {Cinabro}, {DeJongh}, {Depoy}, {Doi}, {Garnavich}, {Hogan}, {Jha}, {Konishi},
  {Lampeitl}, {Marshall}, {McGinnis}, {Prieto}, {Riess}, {Richmond},
  {Schneider}, {Smith}, {Takanashi}, {Tokita}, {van der Heyden}, {Yasuda},
  {Zheng}, {Barentine}, {Brewington}, {Choi}, {Crotts}, {Dembicky}, {Harvanek},
  {Im}, {Ketzeback}, {Kleinman}, {Krzesi{\'n}ski}, {Long}, {Malanushenko},
  {Malanushenko}, {McMillan}, {Nitta}, {Pan}, {Saurage}, {Snedden}, {Watters},
  {Wheeler}, \& {York}}]{Dilday2008ApJ...682..262D}
{Dilday}, B., {Kessler}, R., {Frieman}, J.~A., {et~al.} 2008, \apj, 682, 262

\bibitem[{{Feindt} {et~al.}(2019){Feindt}, {Nordin}, {Rigault}, {Brinnel},
  {Dhawan}, {Goobar}, \& {Kowalski}}]{2019JCAP...10..005F}
{Feindt}, U., {Nordin}, J., {Rigault}, M., {et~al.} 2019, \jcap, 2019, 005

\bibitem[{{Fremling} {et~al.}(2020){Fremling}, {Miller}, {Sharma}, {Dugas},
  {Perley}, {Taggart}, {Sollerman}, {Goobar}, {Graham}, {Neill}, {Nordin},
  {Rigault}, {Walters}, {Andreoni}, {Bagdasaryan}, {Belicki}, {Cannella},
  {Bellm}, {Cenko}, {De}, {Dekany}, {Frederick}, {Golkhou}, {Graham}, {Helou},
  {Ho}, {Kasliwal}, {Kupfer}, {Laher}, {Mahabal}, {Masci}, {Riddle},
  {Rusholme}, {Schulze}, {Shupe}, {Smith}, {van Velzen}, {Yan}, {Yao},
  {Zhuang}, \& {Kulkarni}}]{2020ApJ...895...32F}
{Fremling}, C., {Miller}, A.~A., {Sharma}, Y., {et~al.} 2020, \apj, 895, 32

\bibitem[{{Frye} {et~al.}(2023){Frye}, {Pascale}, {Cohen}, {Summers}, {Foo},
  {Kamieneski}, {Carleton}, {Jansen}, {Pierel}, {Engesser}, {Chen}, {Austin},
  {Marshall}, {Trussler}, {Meena}, {Leimbach}, {Garuda}, {Honor}, {Furtak},
  {Strolger}, {Windhorst}, {Koekemoer}, {Zitrin}, {Diego}, {Kelly}, {Coe},
  {Conselice}, {Dai}, {D{\^a}Silva}, {Dole}, {Driver}, {Grogin}, {Nonino},
  {Pirzkal}, {Polletta}, {Robotham}, {Rutkowski}, {Ryan}, {Tompkins},
  {Willmer}, {Willner}, {Yan}, \& {Yun}}]{Frye2023TNSAN..96....1F}
{Frye}, B., {Pascale}, M., {Cohen}, S., {et~al.} 2023, Transient Name Server
  AstroNote, 96, 1

\bibitem[{{Gilliland} {et~al.}(1999){Gilliland}, {Nugent}, \&
  {Phillips}}]{GillilNugentPhillipsand1999ApJ...521...30G}
{Gilliland}, R.~L., {Nugent}, P.~E., \& {Phillips}, M.~M. 1999, \apj, 521, 30

\bibitem[{{Goldstein} \& {Nugent}(2017)}]{Goldstein2017ApJ...834L...5G}
{Goldstein}, D.~A. \& {Nugent}, P.~E. 2017, \apjl, 834, L5

\bibitem[{{Goldstein} {et~al.}(2019){Goldstein}, {Nugent}, \&
  {Goobar}}]{Goldstein2019ApJS..243....6G}
{Goldstein}, D.~A., {Nugent}, P.~E., \& {Goobar}, A. 2019, \apjs, 243, 6

\bibitem[{{Goobar} {et~al.}(2017){Goobar}, {Amanullah}, {Kulkarni}, {Nugent},
  {Johansson}, {Steidel}, {Law}, {M{\"o}rtsell}, {Quimby}, {Blagorodnova},
  {Brandeker}, {Cao}, {Cooray}, {Ferretti}, {Fremling}, {Hangard}, {Kasliwal},
  {Kupfer}, {Lunnan}, {Masci}, {Miller}, {Nayyeri}, {Neill}, {Ofek},
  {Papadogiannakis}, {Petrushevska}, {Ravi}, {Sollerman}, {Sullivan}, {Taddia},
  {Walters}, {Wilson}, {Yan}, \& {Yaron}}]{Goobar2017Sci...356..291G}
{Goobar}, A., {Amanullah}, R., {Kulkarni}, S.~R., {et~al.} 2017, Science, 356,
  291

\bibitem[{{Goobar} {et~al.}(2023){Goobar}, {Johansson}, {Schulze}, {Arendse},
  {Carracedo}, {Dhawan}, {M{\"o}rtsell}, {Fremling}, {Yan}, {Perley},
  {Sollerman}, {Joseph}, {Hinds}, {Meynardie}, {Andreoni}, {Bellm}, {Bloom},
  {Collett}, {Drake}, {Graham}, {Kasliwal}, {Kulkarni}, {Lemon}, {Miller},
  {Neill}, {Nordin}, {Pierel}, {Richard}, {Riddle}, {Rigault}, {Rusholme},
  {Sharma}, {Stein}, {Stewart}, {Townsend}, {Vinko}, {Wheeler}, \&
  {Wold}}]{2023NatAs...7.1098G}
{Goobar}, A., {Johansson}, J., {Schulze}, S., {et~al.} 2023, Nature Astronomy,
  7, 1098

\bibitem[{{Goobar} {et~al.}(2002){Goobar}, {M{\"o}rtsell}, {Amanullah}, \&
  {Nugent}}]{2002A&A...393...25G}
{Goobar}, A., {M{\"o}rtsell}, E., {Amanullah}, R., \& {Nugent}, P. 2002, \aap,
  393, 25

\bibitem[{{Graham} {et~al.}(2019){Graham}, {Kulkarni}, {Bellm}, {Adams},
  {Barbarino}, {Blagorodnova}, {Bodewits}, {Bolin}, {Brady}, {Cenko}, {Chang},
  {Coughlin}, {De}, {Eadie}, {Farnham}, {Feindt}, {Franckowiak}, {Fremling},
  {Gezari}, {Ghosh}, {Goldstein}, {Golkhou}, {Goobar}, {Ho}, {Huppenkothen},
  {Ivezi{\'c}}, {Jones}, {Juric}, {Kaplan}, {Kasliwal}, {Kelley}, {Kupfer},
  {Lee}, {Lin}, {Lunnan}, {Mahabal}, {Miller}, {Ngeow}, {Nugent}, {Ofek},
  {Prince}, {Rauch}, {van Roestel}, {Schulze}, {Singer}, {Sollerman}, {Taddia},
  {Yan}, {Ye}, {Yu}, {Barlow}, {Bauer}, {Beck}, {Belicki}, {Biswas}, {Brinnel},
  {Brooke}, {Bue}, {Bulla}, {Burruss}, {Connolly}, {Cromer}, {Cunningham},
  {Dekany}, {Delacroix}, {Desai}, {Duev}, {Feeney}, {Flynn}, {Frederick},
  {Gal-Yam}, {Giomi}, {Groom}, {Hacopians}, {Hale}, {Helou}, {Henning},
  {Hover}, {Hillenbrand}, {Howell}, {Hung}, {Imel}, {Ip}, {Jackson}, {Kaspi},
  {Kaye}, {Kowalski}, {Kramer}, {Kuhn}, {Landry}, {Laher}, {Mao}, {Masci},
  {Monkewitz}, {Murphy}, {Nordin}, {Patterson}, {Penprase}, {Porter},
  {Rebbapragada}, {Reiley}, {Riddle}, {Rigault}, {Rodriguez}, {Rusholme}, {van
  Santen}, {Shupe}, {Smith}, {Soumagnac}, {Stein}, {Surace}, {Szkody}, {Terek},
  {Van Sistine}, {van Velzen}, {Vestrand}, {Walters}, {Ward}, {Zhang}, \&
  {Zolkower}}]{Graham2019}
{Graham}, M.~J., {Kulkarni}, S.~R., {Bellm}, E.~C., {et~al.} 2019, \pasp, 131,
  078001

\bibitem[{{Grillo} {et~al.}(2018){Grillo}, {Rosati}, {Suyu}, {Balestra},
  {Caminha}, {Halkola}, {Kelly}, {Lombardi}, {Mercurio}, {Rodney}, \&
  {Treu}}]{Grillo2018ApJ...860...94G}
{Grillo}, C., {Rosati}, P., {Suyu}, S.~H., {et~al.} 2018, \apj, 860, 94

\bibitem[{{Guy} {et~al.}(2007){Guy}, {Astier}, {Baumont}, {Hardin}, {Pain},
  {Regnault}, {Basa}, {Carlberg}, {Conley}, {Fabbro}, {Fouchez}, {Hook},
  {Howell}, {Perrett}, {Pritchet}, {Rich}, {Sullivan}, {Antilogus}, {Aubourg},
  {Bazin}, {Bronder}, {Filiol}, {Palanque-Delabrouille}, {Ripoche}, \&
  {Ruhlmann-Kleider}}]{Guy2007A&A...466...11G}
{Guy}, J., {Astier}, P., {Baumont}, S., {et~al.} 2007, \aap, 466, 11

\bibitem[{{Hounsell} {et~al.}(2018){Hounsell}, {Scolnic}, {Foley}, {Kessler},
  {Miranda}, {Avelino}, {Bohlin}, {Filippenko}, {Frieman}, {Jha}, {Kelly},
  {Kirshner}, {Mandel}, {Rest}, {Riess}, {Rodney}, \&
  {Strolger}}]{Hounsell2018ApJ...867...23H}
{Hounsell}, R., {Scolnic}, D., {Foley}, R.~J., {et~al.} 2018, \apj, 867, 23

\bibitem[{{Ivezi{\'c}} {et~al.}(2019){Ivezi{\'c}}, {Kahn}, {Tyson}, {Abel},
  {Acosta}, {Allsman}, {Alonso}, {AlSayyad}, {Anderson}, {Andrew}, {Angel},
  {Angeli}, {Ansari}, {Antilogus}, {Araujo}, {Armstrong}, {Arndt}, {Astier},
  {Aubourg}, {Auza}, {Axelrod}, {Bard}, {Barr}, {Barrau}, {Bartlett}, {Bauer},
  {Bauman}, {Baumont}, {Bechtol}, {Bechtol}, {Becker}, {Becla}, {Beldica},
  {Bellavia}, {Bianco}, {Biswas}, {Blanc}, {Blazek}, {Blandford}, {Bloom},
  {Bogart}, {Bond}, {Booth}, {Borgland}, {Borne}, {Bosch}, {Boutigny},
  {Brackett}, {Bradshaw}, {Brandt}, {Brown}, {Bullock}, {Burchat}, {Burke},
  {Cagnoli}, {Calabrese}, {Callahan}, {Callen}, {Carlin}, {Carlson},
  {Chandrasekharan}, {Charles-Emerson}, {Chesley}, {Cheu}, {Chiang}, {Chiang},
  {Chirino}, {Chow}, {Ciardi}, {Claver}, {Cohen-Tanugi}, {Cockrum}, {Coles},
  {Connolly}, {Cook}, {Cooray}, {Covey}, {Cribbs}, {Cui}, {Cutri}, {Daly},
  {Daniel}, {Daruich}, {Daubard}, {Daues}, {Dawson}, {Delgado}, {Dellapenna},
  {de Peyster}, {de Val-Borro}, {Digel}, {Doherty}, {Dubois},
  {Dubois-Felsmann}, {Durech}, {Economou}, {Eifler}, {Eracleous}, {Emmons},
  {Fausti Neto}, {Ferguson}, {Figueroa}, {Fisher-Levine}, {Focke}, {Foss},
  {Frank}, {Freemon}, {Gangler}, {Gawiser}, {Geary}, {Gee}, {Geha}, {Gessner},
  {Gibson}, {Gilmore}, {Glanzman}, {Glick}, {Goldina}, {Goldstein}, {Goodenow},
  {Graham}, {Gressler}, {Gris}, {Guy}, {Guyonnet}, {Haller}, {Harris},
  {Hascall}, {Haupt}, {Hernandez}, {Herrmann}, {Hileman}, {Hoblitt}, {Hodgson},
  {Hogan}, {Howard}, {Huang}, {Huffer}, {Ingraham}, {Innes}, {Jacoby}, {Jain},
  {Jammes}, {Jee}, {Jenness}, {Jernigan}, {Jevremovi{\'c}}, {Johns}, {Johnson},
  {Johnson}, {Jones}, {Juramy-Gilles}, {Juri{\'c}}, {Kalirai}, {Kallivayalil},
  {Kalmbach}, {Kantor}, {Karst}, {Kasliwal}, {Kelly}, {Kessler}, {Kinnison},
  {Kirkby}, {Knox}, {Kotov}, {Krabbendam}, {Krughoff}, {Kub{\'a}nek},
  {Kuczewski}, {Kulkarni}, {Ku}, {Kurita}, {Lage}, {Lambert}, {Lange},
  {Langton}, {Le Guillou}, {Levine}, {Liang}, {Lim}, {Lintott}, {Long},
  {Lopez}, {Lotz}, {Lupton}, {Lust}, {MacArthur}, {Mahabal}, {Mandelbaum},
  {Markiewicz}, {Marsh}, {Marshall}, {Marshall}, {May}, {McKercher}, {McQueen},
  {Meyers}, {Migliore}, {Miller}, {Mills}, {Miraval}, {Moeyens}, {Moolekamp},
  {Monet}, {Moniez}, {Monkewitz}, {Montgomery}, {Morrison}, {Mueller},
  {Muller}, {Mu{\~n}oz Arancibia}, {Neill}, {Newbry}, {Nief}, {Nomerotski},
  {Nordby}, {O'Connor}, {Oliver}, {Olivier}, {Olsen}, {O'Mullane}, {Ortiz},
  {Osier}, {Owen}, {Pain}, {Palecek}, {Parejko}, {Parsons}, {Pease},
  {Peterson}, {Peterson}, {Petravick}, {Libby Petrick}, {Petry},
  {Pierfederici}, {Pietrowicz}, {Pike}, {Pinto}, {Plante}, {Plate}, {Plutchak},
  {Price}, {Prouza}, {Radeka}, {Rajagopal}, {Rasmussen}, {Regnault}, {Reil},
  {Reiss}, {Reuter}, {Ridgway}, {Riot}, {Ritz}, {Robinson}, {Roby}, {Roodman},
  {Rosing}, {Roucelle}, {Rumore}, {Russo}, {Saha}, {Sassolas}, {Schalk},
  {Schellart}, {Schindler}, {Schmidt}, {Schneider}, {Schneider}, {Schoening},
  {Schumacher}, {Schwamb}, {Sebag}, {Selvy}, {Sembroski}, {Seppala}, {Serio},
  {Serrano}, {Shaw}, {Shipsey}, {Sick}, {Silvestri}, {Slater}, {Smith},
  {Smith}, {Sobhani}, {Soldahl}, {Storrie-Lombardi}, {Stover}, {Strauss},
  {Street}, {Stubbs}, {Sullivan}, {Sweeney}, {Swinbank}, {Szalay}, {Takacs},
  {Tether}, {Thaler}, {Thayer}, {Thomas}, {Thornton}, {Thukral}, {Tice},
  {Trilling}, {Turri}, {Van Berg}, {Vanden Berk}, {Vetter}, {Virieux},
  {Vucina}, {Wahl}, {Walkowicz}, {Walsh}, {Walter}, {Wang}, {Wang}, {Warner},
  {Wiecha}, {Willman}, {Winters}, {Wittman}, {Wolff}, {Wood-Vasey}, {Wu},
  {Xin}, {Yoachim}, \& {Zhan}}]{Ivezi2019ApJ...873..111I}
{Ivezi{\'c}}, {\v{Z}}., {Kahn}, S.~M., {Tyson}, J.~A., {et~al.} 2019, \apj,
  873, 111

\bibitem[{{Kaiser} {et~al.}(2010){Kaiser}, {Burgett}, {Chambers}, {Denneau},
  {Heasley}, {Jedicke}, {Magnier}, {Morgan}, {Onaka}, \& {Tonry}}]{Kaiser2010}
{Kaiser}, N., {Burgett}, W., {Chambers}, K., {et~al.} 2010, in SPIE, Vol. 7733,
  Ground-based and Airborne Telescopes III, ed. L.~M. {Stepp}, R.~{Gilmozzi},
  \& H.~J. {Hall}, 77330E

\bibitem[{{Kelly} {et~al.}(2022){Kelly}, {Zitrin}, {Oguri}, {Diego},
  {Williams}, {Broadhurst}, {Chen}, {Koekemoer}, {Pierel}, {Strolger}, \&
  {Treu}}]{Kelly2022TNSAN.169....1K}
{Kelly}, P., {Zitrin}, A., {Oguri}, M., {et~al.} 2022, Transient Name Server
  AstroNote, 169, 1

\bibitem[{{Kelly} {et~al.}(2023){Kelly}, {Rodney}, {Treu}, {Oguri}, {Chen},
  {Zitrin}, {Birrer}, {Bonvin}, {Dessart}, {Diego}, {Filippenko}, {Foley},
  {Gilman}, {Hjorth}, {Jauzac}, {Mandel}, {Millon}, {Pierel}, {Sharon},
  {Thorp}, {Williams}, {Broadhurst}, {Dressler}, {Graur}, {Jha}, {McCully},
  {Postman}, {Schmidt}, {Tucker}, \& {von der
  Linden}}]{Kelly2023Sci...380.1322K}
{Kelly}, P.~L., {Rodney}, S., {Treu}, T., {et~al.} 2023, Science, 380, abh1322

\bibitem[{{Kelly} {et~al.}(2015){Kelly}, {Rodney}, {Treu}, {Foley}, {Brammer},
  {Schmidt}, {Zitrin}, {Sonnenfeld}, {Strolger}, {Graur}, {Filippenko}, {Jha},
  {Riess}, {Bradac}, {Weiner}, {Scolnic}, {Malkan}, {von der Linden}, {Trenti},
  {Hjorth}, {Gavazzi}, {Fontana}, {Merten}, {McCully}, {Jones}, {Postman},
  {Dressler}, {Patel}, {Cenko}, {Graham}, \&
  {Tucker}}]{Kelly2015Sci...347.1123K}
{Kelly}, P.~L., {Rodney}, S.~A., {Treu}, T., {et~al.} 2015, Science, 347, 1123

\bibitem[{{Kessler} {et~al.}(2019){Kessler}, {Narayan}, {Avelino}, {Bachelet},
  {Biswas}, {Brown}, {Chernoff}, {Connolly}, {Dai}, {Daniel}, {Di Stefano},
  {Drout}, {Galbany}, {Gonz{\'a}lez-Gait{\'a}n}, {Graham}, {Hlo{\v{z}}ek},
  {Ishida}, {Guillochon}, {Jha}, {Jones}, {Mandel}, {Muthukrishna}, {O'Grady},
  {Peters}, {Pierel}, {Ponder}, {Pr{\v{s}}a}, {Rodney}, {Villar}, {LSST Dark
  Energy Science Collaboration}, \& {Transient and Variable Stars Science
  Collaboration}}]{Kessler2019PASP..131i4501K}
{Kessler}, R., {Narayan}, G., {Avelino}, A., {et~al.} 2019, \pasp, 131, 094501

\bibitem[{{Kormann} {et~al.}(1994){Kormann}, {Schneider}, \&
  {Bartelmann}}]{Kormann1994A&A...284..285K}
{Kormann}, R., {Schneider}, P., \& {Bartelmann}, M. 1994, \aap, 284, 285

\bibitem[{{Kulkarni}(2013)}]{Kulkarni2013ATel.4807....1K}
{Kulkarni}, S.~R. 2013, The Astronomer's Telegram, 4807, 1

\bibitem[{{Levan} {et~al.}(2005){Levan}, {Nugent}, {Fruchter}, {Burud},
  {Branch}, {Rhoads}, {Castro-Tirado}, {Gorosabel}, {Castro Cer{\'o}n},
  {Thorsett}, {Kouveliotou}, {Golenetskii}, {Fynbo}, {Garnavich}, {Holland},
  {Hjorth}, {M{\o}ller}, {Pian}, {Tanvir}, {Ulanov}, {Wijers}, \&
  {Woosley}}]{LevanNugent2005ApJ...624..880L}
{Levan}, A., {Nugent}, P., {Fruchter}, A., {et~al.} 2005, \apj, 624, 880

\bibitem[{{Madau} \& {Dickinson}(2014)}]{Madau2014ARA&A..52..415M}
{Madau}, P. \& {Dickinson}, M. 2014, \araa, 52, 415

\bibitem[{{Nicolas} {et~al.}(2021){Nicolas}, {Rigault}, {Copin}, {Graziani},
  {Aldering}, {Briday}, {Kim}, {Nordin}, {Perlmutter}, \&
  {Smith}}]{Nicolas2021A&A...649A..74N}
{Nicolas}, N., {Rigault}, M., {Copin}, Y., {et~al.} 2021, \aap, 649, A74

\bibitem[{{Oguri}(2019)}]{Oguri2019RPPh...82l6901O}
{Oguri}, M. 2019, Reports on Progress in Physics, 82, 126901

\bibitem[{{Oguri} \& {Marshall}(2010)}]{Oguri2010MNRAS.405.2579O}
{Oguri}, M. \& {Marshall}, P.~J. 2010, \mnras, 405, 2579

\bibitem[{{Perley} {et~al.}(2020){Perley}, {Fremling}, {Sollerman}, {Miller},
  {Dahiwale}, {Sharma}, {Bellm}, {Biswas}, {Brink}, {Bruch}, {De}, {Dekany},
  {Drake}, {Duev}, {Filippenko}, {Gal-Yam}, {Goobar}, {Graham}, {Graham}, {Ho},
  {Irani}, {Kasliwal}, {Kim}, {Kulkarni}, {Mahabal}, {Masci}, {Modak}, {Neill},
  {Nordin}, {Riddle}, {Soumagnac}, {Strotjohann}, {Schulze}, {Taggart},
  {Tzanidakis}, {Walters}, \& {Yan}}]{Perley2020ApJ...904...35P}
{Perley}, D.~A., {Fremling}, C., {Sollerman}, J., {et~al.} 2020, \apj, 904, 35

\bibitem[{{Pierel} {et~al.}(2023)}]{Pierel_2023_Encore}
{Pierel}, J. {et~al.} 2023, {Lensed Supernova Encore at $z=2$! The First Galaxy
  to Host Two Multiply-Imaged Supernovae}, JWST Proposal. Cycle 2, ID. \#6549

\bibitem[{{Pierel} {et~al.}(2021){Pierel}, {Rodney}, {Vernardos}, {Oguri},
  {Kessler}, \& {Anguita}}]{Pierel2021ApJ...908..190P}
{Pierel}, J.~D.~R., {Rodney}, S., {Vernardos}, G., {et~al.} 2021, \apj, 908,
  190

\bibitem[{{Planck Collaboration} {et~al.}(2020){Planck Collaboration},
  {Aghanim}, {Akrami}, {Ashdown}, {Aumont}, {Baccigalupi}, {Ballardini},
  {Banday}, {Barreiro}, {Bartolo}, {Basak}, {Battye}, {Benabed}, {Bernard},
  {Bersanelli}, {Bielewicz}, {Bock}, {Bond}, {Borrill}, {Bouchet}, {Boulanger},
  {Bucher}, {Burigana}, {Butler}, {Calabrese}, {Cardoso}, {Carron},
  {Challinor}, {Chiang}, {Chluba}, {Colombo}, {Combet}, {Contreras}, {Crill},
  {Cuttaia}, {de Bernardis}, {de Zotti}, {Delabrouille}, {Delouis}, {Di
  Valentino}, {Diego}, {Dor{\'e}}, {Douspis}, {Ducout}, {Dupac}, {Dusini},
  {Efstathiou}, {Elsner}, {En{\ss}lin}, {Eriksen}, {Fantaye}, {Farhang},
  {Fergusson}, {Fernandez-Cobos}, {Finelli}, {Forastieri}, {Frailis},
  {Fraisse}, {Franceschi}, {Frolov}, {Galeotta}, {Galli}, {Ganga},
  {G{\'e}nova-Santos}, {Gerbino}, {Ghosh}, {Gonz{\'a}lez-Nuevo}, {G{\'o}rski},
  {Gratton}, {Gruppuso}, {Gudmundsson}, {Hamann}, {Handley}, {Hansen},
  {Herranz}, {Hildebrandt}, {Hivon}, {Huang}, {Jaffe}, {Jones}, {Karakci},
  {Keih{\"a}nen}, {Keskitalo}, {Kiiveri}, {Kim}, {Kisner}, {Knox},
  {Krachmalnicoff}, {Kunz}, {Kurki-Suonio}, {Lagache}, {Lamarre}, {Lasenby},
  {Lattanzi}, {Lawrence}, {Le Jeune}, {Lemos}, {Lesgourgues}, {Levrier},
  {Lewis}, {Liguori}, {Lilje}, {Lilley}, {Lindholm}, {L{\'o}pez-Caniego},
  {Lubin}, {Ma}, {Mac{\'\i}as-P{\'e}rez}, {Maggio}, {Maino}, {Mandolesi},
  {Mangilli}, {Marcos-Caballero}, {Maris}, {Martin}, {Martinelli},
  {Mart{\'\i}nez-Gonz{\'a}lez}, {Matarrese}, {Mauri}, {McEwen}, {Meinhold},
  {Melchiorri}, {Mennella}, {Migliaccio}, {Millea}, {Mitra},
  {Miville-Desch{\^e}nes}, {Molinari}, {Montier}, {Morgante}, {Moss}, {Natoli},
  {N{\o}rgaard-Nielsen}, {Pagano}, {Paoletti}, {Partridge}, {Patanchon},
  {Peiris}, {Perrotta}, {Pettorino}, {Piacentini}, {Polastri}, {Polenta},
  {Puget}, {Rachen}, {Reinecke}, {Remazeilles}, {Renzi}, {Rocha}, {Rosset},
  {Roudier}, {Rubi{\~n}o-Mart{\'\i}n}, {Ruiz-Granados}, {Salvati}, {Sandri},
  {Savelainen}, {Scott}, {Shellard}, {Sirignano}, {Sirri}, {Spencer},
  {Sunyaev}, {Suur-Uski}, {Tauber}, {Tavagnacco}, {Tenti}, {Toffolatti},
  {Tomasi}, {Trombetti}, {Valenziano}, {Valiviita}, {Van Tent}, {Vibert},
  {Vielva}, {Villa}, {Vittorio}, {Wandelt}, {Wehus}, {White}, {White},
  {Zacchei}, \& {Zonca}}]{Planck2020A&A...641A...6P}
{Planck Collaboration}, {Aghanim}, N., {Akrami}, Y., {et~al.} 2020, \aap, 641,
  A6

\bibitem[{{Refsdal}(1964)}]{Refsdal1964MNRAS.128..307R}
{Refsdal}, S. 1964, \mnras, 128, 307

\bibitem[{{Rodney} {et~al.}(2021){Rodney}, {Brammer}, {Pierel}, {Richard},
  {Toft}, {O'Connor}, {Akhshik}, \& {Whitaker}}]{Rodney2021NatAs...5.1118R}
{Rodney}, S.~A., {Brammer}, G.~B., {Pierel}, J. D.~R., {et~al.} 2021, Nature
  Astronomy, 5, 1118

\bibitem[{{Sainz de Murieta} {et~al.}(2023){Sainz de Murieta}, {Collett},
  {Magee}, {Weisenbach}, {Krawczyk}, \& {Enzi}}]{Sainz2023MNRAS.526.4296S}
{Sainz de Murieta}, A., {Collett}, T.~E., {Magee}, M.~R., {et~al.} 2023,
  \mnras, 526, 4296

\bibitem[{{Sako} {et~al.}(2011){Sako}, {Bassett}, {Connolly}, {Dilday},
  {Cambell}, {Frieman}, {Gladney}, {Kessler}, {Lampeitl}, {Marriner}, {Miquel},
  {Nichol}, {Schneider}, {Smith}, \& {Sollerman}}]{Sako2011ApJ...738..162S}
{Sako}, M., {Bassett}, B., {Connolly}, B., {et~al.} 2011, \apj, 738, 162

\bibitem[{{Schlegel} {et~al.}(1998){Schlegel}, {Finkbeiner}, \&
  {Davis}}]{Schlegel1998ApJ...500..525S}
{Schlegel}, D.~J., {Finkbeiner}, D.~P., \& {Davis}, M. 1998, \apj, 500, 525

\bibitem[{{Shappee} {et~al.}(2014){Shappee}, {Prieto}, {Grupe}, {Kochanek},
  {Stanek}, {De Rosa}, {Mathur}, {Zu}, {Peterson}, {Pogge}, {Komossa}, {Im},
  {Jencson}, {Holoien}, {Basu}, {Beacom}, {Szczygie{\l}}, {Brimacombe},
  {Adams}, {Campillay}, {Choi}, {Contreras}, {Dietrich}, {Dubberley},
  {Elphick}, {Foale}, {Giustini}, {Gonzalez}, {Hawkins}, {Howell}, {Hsiao},
  {Koss}, {Leighly}, {Morrell}, {Mudd}, {Mullins}, {Nugent}, {Parrent},
  {Phillips}, {Pojmanski}, {Rosing}, {Ross}, {Sand}, {Terndrup}, {Valenti},
  {Walker}, \& {Yoon}}]{Shappee2014}
{Shappee}, B.~J., {Prieto}, J.~L., {Grupe}, D., {et~al.} 2014, \apj, 788, 48

\bibitem[{{Sheth} {et~al.}(2003){Sheth}, {Bernardi}, {Schechter}, {Burles},
  {Eisenstein}, {Finkbeiner}, {Frieman}, {Lupton}, {Schlegel}, {Subbarao},
  {Shimasaku}, {Bahcall}, {Brinkmann}, \&
  {Ivezi{\'c}}}]{Sheth2003ApJ...594..225S}
{Sheth}, R.~K., {Bernardi}, M., {Schechter}, P.~L., {et~al.} 2003, \apj, 594,
  225

\bibitem[{{Stanishev} {et~al.}(2018){Stanishev}, {Goobar}, {Amanullah},
  {Bassett}, {Fantaye}, {Garnavich}, {Hlozek}, {Nordin}, {Okouma},
  {{\"O}stman}, {Sako}, {Scalzo}, \& {Smith}}]{Stanishev2018}
{Stanishev}, V., {Goobar}, A., {Amanullah}, R., {et~al.} 2018, \aap, 615, A45

\bibitem[{{Suyu} {et~al.}(2024){Suyu}, {Goobar}, {Collett}, {More}, \&
  {Vernardos}}]{2024SSRv..220...13S}
{Suyu}, S.~H., {Goobar}, A., {Collett}, T., {More}, A., \& {Vernardos}, G.
  2024, \ssr, 220, 13

\bibitem[{{Tonry} {et~al.}(2018){Tonry}, {Denneau}, {Heinze}, {Stalder},
  {Smith}, {Smartt}, {Stubbs}, {Weiland }, \& {Rest}}]{Tonry2018}
{Tonry}, J.~L., {Denneau}, L., {Heinze}, A.~N., {et~al.} 2018, \pasp, 130,
  064505

\bibitem[{{Treu} \& {Marshall}(2016)}]{2016A&ARv..24...11T}
{Treu}, T. \& {Marshall}, P.~J. 2016, \aapr, 24, 11

\bibitem[{{Tripp}(1998)}]{Tripp1998A&A...331..815T}
{Tripp}, R. 1998, \aap, 331, 815

\bibitem[{{Wojtak} {et~al.}(2019){Wojtak}, {Hjorth}, \&
  {Gall}}]{Wojtak2019MNRAS.487.3342W}
{Wojtak}, R., {Hjorth}, J., \& {Gall}, C. 2019, \mnras, 487, 3342

\bibitem[{{Wong} {et~al.}(2011){Wong}, {Keeton}, {Williams}, {Momcheva}, \&
  {Zabludoff}}]{Wong2011ApJ...726...84W}
{Wong}, K.~C., {Keeton}, C.~R., {Williams}, K.~A., {Momcheva}, I.~G., \&
  {Zabludoff}, A.~I. 2011, \apj, 726, 84

\end{thebibliography}

\begin{appendix}

\section{Photometric error impact on inferred brightness}\label{sec:M_bias_phzerr}

\begin{figure}
    \centering
    \includegraphics[width=\linewidth]{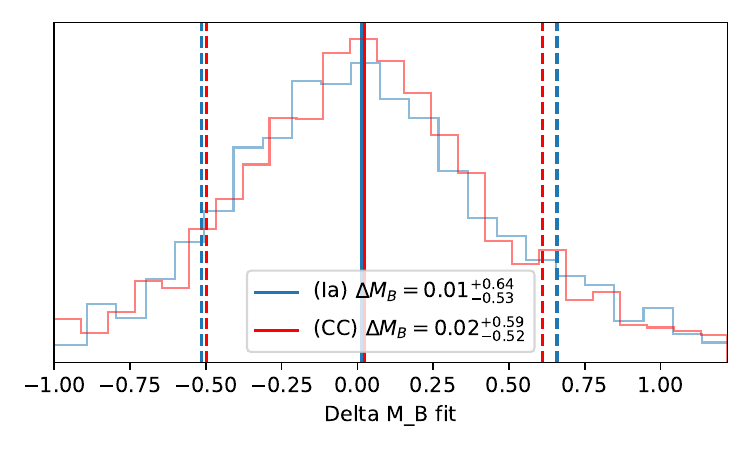}
    \caption{We show the bias $\rm \Delta M_B$ due to the photometric error. The red histogram shows for CC supernovae and the blue for Ia supernovae. Shown median with 68\% CI. There is no signs on different biases on the infered $M_B$ for CCSN and Ia. }
    \label{fig:d_mb}
\end{figure}

We investigate the effect of the photometric redshift error in the inference of the intrinsic absolute magnitude of the supernovae. We apply a 15\% standard deviation from the true redshift of the lens galaxy. 
We investigate the bias on $M_{\rm B, zl}$ obtained from the SALT2 fit with and without the photometric error
\begin{equation}
    \Delta M_{\rm B, zl} = M_{\rm B, zl,phot} - M_{\rm B, zl},
\end{equation} finding $\Delta M_{\rm B, zl}=0.02^{+0.68}_{-0.58}$ mag (see Fig.~\ref{fig:d_mb}). 
We note that the bias in the inferred $M_{\rm B, zl}$ is null in average, but in individual cases it can bias up to $\sim\pm1$ magnitude, which can induce false candidates and exclude true candidates. We find that $\sim 6.7$\% of the events would be lost from the photometric error bias, meaning that we get $<-19.5$ with the true redshift but $\geq-19.5$ with the photometric error. Inversely, we find that $\sim 5.6$\% would be a false candidate, as they would only pass the brightness cut with the photometric error applied. In the case of glSNe\-CC we get 4.9\% losses and 4.7\% false passes. Nevertheless, the symmetry in this bias indicates that the error bias in $M_{\rm B, zl}$ is not affecting the rate estimates. See Fig.~\ref{fig:d_mb} for a visual distribution of this bias.

\end{appendix}

\end{document}